\documentclass[final,3p,times,twocolumn,authoryear]{elsarticle}
\graphicspath{ {./figs/} }
\usepackage{hyperref}
\usepackage{apalike}
\usepackage{graphicx}%
\usepackage{multirow}%
\usepackage{amsmath,amssymb,amsfonts}%
\usepackage{amsthm}%
\usepackage{mathrsfs}%
\usepackage{xcolor}%
\usepackage{textcomp}%
\usepackage{manyfoot}%
\usepackage{booktabs}%
\usepackage{algorithm}%
\usepackage{algorithmicx}%
\usepackage{algpseudocode}%
\usepackage{listings}%
\usepackage{placeins,float}
\usepackage{multicol}
\usepackage{bm,csquotes}
\usepackage{enumerate}
\usepackage{subcaption}
\usepackage{caption}
\usepackage{array}
\usepackage{booktabs}

\journal{Astronomy and Computing}

\biboptions{authoryear}

\begin{document}
\begin{frontmatter}










\title{Decoding the Radio Sky: Bayesian Ensemble Learning and SVD-Based Feature Extraction for Automated Radio Galaxy Classification}

\author[label1]{Theophilus Ansah-Narh\corref{cor1}}
\ead{theophilus.ansah-narh@gaec.gov.gh}

\author[label2]{Jordan Lontsi Tedongmo}
\author[label1]{Joseph Bremang Tandoh}
\author[label3]{Nia Imara}
\author[label4]{Ezekiel Nii Noye Nortey}

\cortext[cor1]{Corresponding author.}
\address[label1]{Ghana Space Science and Technology Institute, Ghana Atomic Energy Commission, P. O. Box LG 80, Legon-Accra, Ghana.}
\address[label2]{African Institute for Mathematical Sciences Ghana, P. O. Box LG DTD 20046, Legon-Accra, Ghana}
\address[label3]{Astronomy and Astrophysics Department,  University of California, Santa Cruz, CA, USA.}
\address[label4]{Department of Statistics and Actuarial Science, University of Ghana, 
P. O. Box LG 115, Legon-Accra, Ghana.}

\begin{abstract}
The classification of radio galaxies is central to understanding galaxy evolution, active galactic nuclei dynamics, and the large-scale structure of the universe. However, traditional manual techniques are inadequate for processing the massive, heterogeneous datasets generated by modern radio surveys. In this study, we present a probabilistic machine learning framework that integrates Singular Value Decomposition (SVD) for feature extraction with Bayesian ensemble learning to achieve robust, scalable radio galaxy classification.
The SVD approach effectively reduces dimensionality while preserving key morphological structures, enabling efficient representation of galaxy features. To mitigate class imbalance and avoid the introduction of artefacts, we incorporate a Local Neighbourhood Encoding strategy tailored to the astrophysical distribution of galaxy types. The resulting features are used to train and optimise several baseline classifiers: Logistic Regression, Support Vector Machines, LightGBM, and Multi-Layer Perceptrons within bagging, boosting, and stacking ensembles governed by a Bayesian weighting scheme.
Our results demonstrate that Bayesian ensembles outperform their traditional counterparts across all metrics, with the Bayesian stacking model achieving a classification accuracy of \(99.0\%\) and an F1-score of \(0.99\) across Compact, Bent, Fanaroff-Riley Type I (FR-I), and Type II (FR-II) sources.
Interpretability is enhanced through SHAP analysis, which highlights the principal components most associated with morphological distinctions. Beyond improving classification performance, our framework facilitates uncertainty quantification, paving the way for more reliable integration into next-generation survey pipelines. This work contributes a reproducible and interpretable methodology for automated galaxy classification in the era of data-intensive radio astronomy.
\end{abstract}

\begin{keyword}
Radio Galaxy Classification \sep Singular Value Decomposition \sep Class Imbalance Correction \sep Machine Learning in Astronomy \sep Bayesian Ensemble Learning \sep SHAP Interpretability
\end{keyword}

\end{frontmatter}

\section{Introduction}\label{sec:intro}

The classification of radio galaxies plays a crucial role in advancing our understanding of galaxy evolution, the physics of active galactic nuclei (AGN), and the larger cosmic structure
\citep{fajardo2024mapping,mucesh2024galaxy,tudorache2024evolution}.
These include Fanaroff-Riley Class I and II (FR-I and FR-II) galaxies, compact steep-spectrum sources, bent-tail galaxies, and gigahertz peaked-spectrum galaxies.
Radio galaxies, which emit significant amounts of electromagnetic radiation in the radio frequency range, provide unique insights into the underlying mechanisms of supermassive black holes and the complex dynamics within galaxy clusters \citep{padovani2017active,STACY2003397}. 
This emission is most effectively detected and studied using radio interferometers.
By leveraging the principles of interferometry, these systems enhance both the resolution and sensitivity of observations, enabling astronomers to capture intricate details of radio emissions from galaxies with exceptional clarity \citep{hardcastle2020radio}.

Traditional methods of manual classification played a crucial role in the early development of astronomy but have become inadequate for larger-scale studies, particularly with the enormous datasets resulting from contemporary astronomical surveys. 
These methods often involve astronomers examining images or spectral data by eye and classifying galaxies according to set criteria--an approach that is well-suited for small or familiar datasets but becomes exceedingly labour-intensive and impracticable at a larger scale.
Initiatives such as LOFAR\footnote{\url{https://www.astron.nl/telescopes/lofar/}} \citep{drake2024lofar,cochrane2023lofar,dabhade2020giant} and the SKA\footnote{\url{https://www.skao.int/en}} \citep{vacca2024filaments,wang2024square} produce massive volumes of data, far exceeding the capacity of manual classification methods to keep pace.
Beyond the issue of scale, manual approaches are also prone to errors, inconsistencies, and subjectivity, particularly when dealing with ambiguous features or low-quality observations \citep{beck2018integrating,kuminski2014combining}. 
Moreover, the presence of class imbalance and observational noise further hampers the effectiveness of manual methods, especially when atypical sources are scarce or when important morphological details are obscured \citep{johnson2022survey}. These obstacles underscore the necessity for more robust, scalable, and automated classification solutions.

Machine learning (ML) classification offers a viable alternative to traditional manual techniques, allowing for the handling of extensive datasets with enhanced speed, accuracy, and consistency while minimising human bias and boosting reproducibility.
Despite advancements in integrating ML into astronomy, current methods still encounter significant challenges.
Many models find it difficult to manage issues associated with radio galaxy datasets such as high dimensionality, observational noise, and class imbalance.

In recent studies focused on galaxy classification \citep{zhang2022automatic,scaife2021fanaroff,becker2021cnn,tang2019transfer,alhassan2018first,lukic2018radio,aniyan2017classifying}, deep learning approaches, notably Convolutional Neural Networks (CNNs), have emerged as preferred methods. However, their application has largely been confined to specific neural network architectures, and these models often lack techniques to reliably assess uncertainty.
Moreover, deep learning strategies require substantial amounts of high-quality labelled data, which are not always available in the field of radio astronomy. These models also tend to be computationally demanding, creating difficulties for real-time scrutiny and processing of the vast data derived from contemporary radio surveys. Additionally, adapting to changing classification standards or capturing the intricate and frequently subtle morphological diversity of radio galaxies remains a challenge. These limitations collectively highlight the persistent necessity for more adaptable, interpretable, and data-efficient classification methods.

Feature extraction remains a critical bottleneck in automated classification. Classical methods such as principal component analysis (PCA) and wavelet transforms have been successfully applied to radio galaxy images, yet they may not fully capture the diversity of morphological structures present in complex, high-dimensional data. Matrix decomposition approaches like Singular Value Decomposition (SVD), which underpin PCA, offer a mathematically robust framework for dimensionality reduction. However, their direct application as standalone feature extractors in radio galaxy classification has received comparatively limited attention in the literature \citep{golub2013matrix}.

A further gap in the literature concerns the limited adoption of probabilistic ensemble learning methods in astronomy. Ensemble techniques such as bagging, boosting, and stacking combine multiple models to improve predictive accuracy and reduce variance \citep{kunapuli2023ensemble,zhou2012ensemble}, yet they remain underutilised in radio galaxy classification, particularly in contexts where uncertainty quantification is crucial.

This work introduces a framework for automated radio galaxy classification that uniquely integrates SVD-based feature extraction with uncertainty-aware Bayesian ensemble learning.
While prior studies have predominantly focused on CNNs, our approach offers a complementary paradigm, enhancing interpretability and robustness to class imbalance and noise.
The primary objective is to develop a robust method that can effectively address the challenges posed by large, noisy datasets and class imbalances while improving classification accuracy. By integrating probabilistic ensemble learning methods with advanced feature extraction techniques, we aim to establish a framework that offers superior performance over traditional approaches, enabling the automated classification of radio galaxies on a large scale.
Hence, the objectives of this study are to perform the following tasks: 

\begin{enumerate}[i.]
    \item Formulate an ensemble learning framework that combines boosting, bagging, stacking, and averaging under a Bayesian paradigm to enhance robustness and quantify uncertainty.
    
    \item Apply SVD for dimensionality reduction, enabling efficient feature extraction while preserving key morphological traits of radio galaxies.

    \item Integrate Local Neighbourhood Encoding (LNE) to address class imbalance by generating representative synthetic samples without introducing artefacts.

    \item Use Particle Swarm Optimisation (PSO) to fine-tune hyperparameters of base models, thereby maximising ensemble performance.

    \item Interpret model predictions using SHAP\footnote{SHapley Additive exPlanations} values to identify the most influential principal components and enhance explainability.

    \item Validate the proposed framework by benchmarking its performance against traditional ML methods across multiple metrics.
\end{enumerate}

The paper is structured as follows: Section \ref{sec:Methodoly} describes the dataset of radio galaxies sourced from surveys such as FIRST and NVSS, categorised into Compact, Bent, FR-I, and FR-II types. It also details the use of SVD for feature extraction, enabling dimensionality reduction while preserving key morphological features. To address class imbalance, the LNE method is employed, ensuring balanced datasets without introducing artefacts. Several classifiers, including LightGBM and multilayer perceptrons, are optimised using PSO for enhanced performance. Model evaluation for multi-class classification is presented in Section \ref{sec:pmetric}.
Results and discussions are provided in Section \ref{sec:RnD}, with the conclusion summarised in Section \ref{sec:conc}.

\section{Materials and Methods} \label{sec:Methodoly}

\subsection{Radio Galaxy Dataset}\label{sec:dataset}

The dataset utilised in this study is primarily based on the FIRST\footnote{Faint Images of the Radio Sky at Twenty Centimetres} survey, augmented by complementary data from the NVSS and LoTSS surveys. The FIRST survey, conducted with the Very Large Array (VLA) in its B-configuration, operates at $1.4$ GHz and achieves an angular resolution of approximately $5^{\prime\prime}$ \citep{Becker1995}. Covering nearly $10,000$ square degrees, it provides high-resolution and high-sensitivity radio images with a detection threshold of about 1 mJy, cataloguing over $946,000$ sources. These characteristics make FIRST exceptionally well-suited for detailed morphological analyses of radio galaxies.

To enhance spectral and spatial coverage, data from the NVSS (NRAO VLA Sky Survey) and LoTSS (LOFAR Two-metre Sky Survey) were incorporated. NVSS, conducted with the VLA in D-configuration, delivers lower-resolution ($45^{\prime\prime}$) observations over a broader sky area, beneficial for identifying extended diffuse emission. In contrast, LoTSS, operating in the 120--168 MHz frequency range, provides insights into low-frequency structures, with an angular resolution of about $6^{\prime\prime}$ under optimal conditions. Together, these surveys offer complementary perspectives on radio morphology across varying frequencies and spatial scales.

For consistency across datasets, a cross-matching procedure was implemented using a positional tolerance of $5^{\prime\prime}$. This radius reflects a compromise between the precise astrometry of FIRST and the broader positional uncertainties of NVSS and LoTSS. Duplicate entries were reconciled by retaining the highest-resolution instance unless multi-frequency context warranted inclusion of lower-resolution data. For regions overlapping between FIRST and LoTSS, FIRST images were convolved to match the LoTSS resolution, allowing consistent morphological comparison. NVSS sources were used primarily to supplement extended emission but were excluded from morphology-based classification due to resolution constraints. To ensure reliability, only sources with signal-to-noise ratios above 10 and flux densities exceeding 5 mJy in at least one band were included.

The final sample was constructed from sky regions jointly covered by FIRST, NVSS, and LoTSS, with preference given to fields also observed by the SDSS to facilitate optical cross-identification. Source selection was based on discernible morphological features in the FIRST survey, with manual classification into four primary categories: Fanaroff–Riley Type I (FR-I), Fanaroff–Riley Type II (FR-II), Compact, and Bent sources. Classification was performed by domain experts and cross-validated with existing catalogues where applicable. Of the initial pool of approximately 3,500 candidates, 2,156 sources were retained based on criteria including morphological clarity, multi-survey completeness, and minimal contamination from imaging artefacts. Class representation was balanced to reflect natural prevalence while maintaining statistical robustness for downstream analysis. 
The dataset includes the following distributions:

\begin{enumerate}[i.]
    \item FR-I sources (495 samples) exhibit diffuse, low-surface-brightness lobes with peak emission near the host galaxy core \citep{fanaroff1974morphology}. Their jets typically decelerate and fade with distance, producing asymmetrical structures shaped by environmental interactions \citep{laing2007jet,ghisellini2001dividing}. These characteristics often indicate less energetic central engines and significant influence from the intergalactic medium (IGM) \citep{best2005host,feretti2002radio}.

    \item FR-II sources (923 samples) are distinguished by their edge-brightened lobes and prominent hotspots, often symmetrically positioned at the termini of narrow, collimated jets. These features suggest more powerful jet dynamics and efficient energy transfer from the central engine to the lobes \citep{fanaroff1974morphology}. FR-II structures are generally associated with higher radio luminosities and more massive host galaxies, indicating a strong correlation between source morphology and intrinsic power \citep{ghisellini2001dividing}.

    \item Compact sources (391 samples) appear unresolved or barely resolved in FIRST images, consistent with point-like emissions. These are often identified as quasars or active galactic nuclei (AGN), where compactness may be due to either small intrinsic size or projection effects \citep{blandford2019relativistic}. Many exhibit characteristics of flat-spectrum radio quasars or BL Lac objects, with relativistic jets oriented close to the line of sight \citep{ghisellini2011transition,lister2009mojave}.

    \item Bent sources (347 samples) display pronounced curvature in their jets and lobes, typically indicative of motion through a dense medium such as a galaxy cluster. These distortions, often described as wide or narrow angle tails, result from ram pressure or environmental asymmetries \citep{freeland2008bent,miley1980structure,blanton2014extragalactic}. Their prevalence in cluster environments underscores their utility as a tracer of large-scale structure and intra-cluster dynamics \citep{mguda2021bent,hardcastle2019radio,sabater2019lotss}.
\end{enumerate}

The selected sources span a broad range of physical properties.
Redshift information, either spectroscopic or photometric was available for approximately $82\%$ of the dataset, predominantly sourced from SDSS DR17. Redshifts range from $z \sim 0.01$ to $z \sim 2.3$, with a median of $0.47$.
Angular extents vary from unresolved (typically $<5^{\prime\prime}$ in the highest-resolution datasets) to over $2^{\prime}$, with a median size of $\sim18^{\prime\prime}$. These measurements reflect a combination of data from the FIRST ($\sim5^{\prime\prime}$ resolution), LoTSS ($\sim6^{\prime\prime}$ to $20^{\prime\prime}$), and NVSS ($\sim45^{\prime\prime}$) surveys, each with differing angular resolutions. Where multiple measurements were available, the angular size from the highest-resolution match was adopted. Optical magnitudes in the $r$\footnote{Covers red visible light (centred around $\rm \sim620\,nm$).} and $i$\footnote{Covers near-infrared light (centred around $\rm \sim750$–$770\,nm$).} bands ranged between $17$ and $22$ mag.
Although ancillary data were not directly employed in the morphological classification, they provide valuable context for interpreting radio galaxy properties and will support future studies on AGN evolution. 
Some incompleteness remains due to faint optical counterparts or positional mismatches, particularly at higher redshifts.

\begin{figure*}
\begin{minipage}[H]{\linewidth}
\centering
\includegraphics[width=\textwidth]{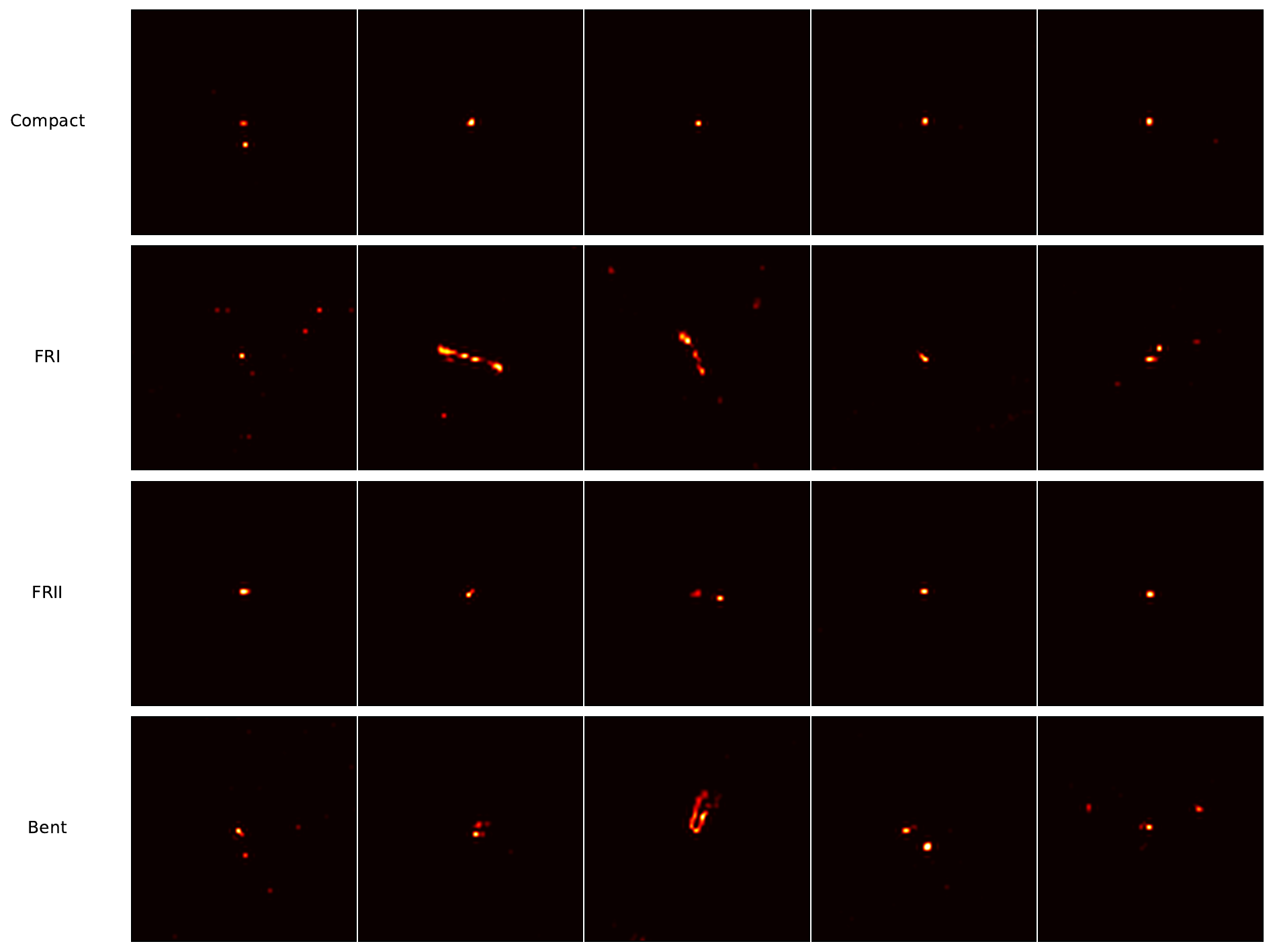} 
\end{minipage}
\caption{A visual representation of the four main types of radio galaxies: Compact, FRI, FRII, and Bent. Each row showcases examples of galaxies with distinct radio morphologies, revealing clues about their activity and evolution.
\textbf{Note:} Each $128 \times 128$ pixel stamp corresponds to an angular field of view of approximately $230^{\prime\prime} \times 230^{\prime\prime}$ (i.e., $\approx 3.8^{\prime} \times 3.8^{\prime}$), based on the FIRST survey pixel scale of $1.8^{\prime\prime}$ per pixel.
}\label{fig:_radio_galaxies_}
\end{figure*}

 Fig.~\ref{fig:_radio_galaxies_} illustrates the four distinct types of radio galaxies considered in this work: Compact, FR-I, FR-II, and Bent. Compact galaxies (top row) exhibit small, concentrated radio emissions around the central source. FR-I galaxies (second row) display extended radio emissions with two-sided jets and diffuse lobes. FR-II galaxies (third row) are characterised by large, well-defined radio lobes and hotspots. Bent galaxies (bottom row) feature distorted or bent radio jets, indicative of interactions with the surrounding environment. These diverse morphologies reflect varying levels of activity and evolutionary stages in these powerful extragalactic objects.

\subsection{Preprocessing} \label{sec:ga-ann}

Preprocessing is a critical step in preparing radio galaxy images for automated classification, ensuring that the data are consistent and optimised for subsequent feature extraction and model training. The heterogeneity in radio galaxy images, in terms of resolution, intensity distribution, and noise levels, necessitates careful preprocessing to mitigate variability and enhance model performance. 

One of the primary preprocessing tasks involves resizing all images to uniform spatial dimensions of \(128 \times 128\) pixels.

This image size strikes a balance between retaining sufficient structural information for classification and reducing computational overhead during feature extraction and model training. 
Let \(I(x, y)\) denote the original intensity at pixel coordinates \((x, y)\) in an image of size \(M \times N\). 
Resizing can be formulated as finding the interpolated intensity \(I'(x', y')\) in the target resolution:
\begin{equation}\label{eq:interp}
I'(x', y') = \text{interpolation}(I, \frac{x'}{128} \cdot M, \frac{y'}{128} \cdot N),
\end{equation}
where \(\text{interpolation}\) may employ methods such as bilinear or bicubic interpolation. Uniform resizing ensures that structural features across the dataset are comparable, aiding in consistent SVD-based feature extraction and reducing the risk of model bias toward larger or smaller objects.

After resizing, pixel intensity normalisation is applied to standardise the image data's dynamic range. This process adjusts pixel values to fall within a [0, 1] range, ensuring numerical stability in computations while preserving the relative intensity details. Let \(I_{\text{min}}\) and \(I_{\text{max}}\) denote the lowest and highest pixel intensities in an image. The normalisation is defined as:
\begin{equation}\label{eq:norm}
I_{\text{norm}}(x, y) = \frac{I(x, y) - I_{\text{min}}}{I_{\text{max}} - I_{\text{min}}}.
\end{equation}
This operation ensures that all images contribute equally to the training process, preventing biases arising from variations in intensity ranges across different samples. Normalisation has been demonstrated to improve the convergence of machine learning models by eliminating large numerical disparities in input data \citep{huang2023normalization,wu2018l1}.

\subsection{Feature Extraction} \label{sec:fextract}
The input dataset \( \mathbf{X} \) is a four-dimensional array of shape \( (N, H, W, C) \), where  \( N \): Number of samples (radio galaxies), \( H, W \): Height and width of the spatial grid (image dimensions), \( C \): Number of feature channels (e.g., intensity measurements). 
To prepare for matrix decomposition, each feature channel \( c \) is reshaped into a 2D matrix:
\begin{equation}\label{eq:xc}
\mathbf{X}_c \in \mathbb{R}^{N \times M}, \quad \text{where } M = H \times W.
\end{equation}

\noindent Given the observational nature of radio astronomy datasets, missing data is inevitable. To address this, the extraction method begins by imputing missing values using the Multiple Imputation by Chained Equations (MICE) approach, a statistical methodology designed for handling incomplete datasets \citep{little2019statistical}.  
 Missing entries in \( \mathbf{X}_c \) are denoted by \( \mathbf{\kappa} \), such that \( \mathbf{X}_c = X_{\text{observed}} \cup \mathbf{\kappa} \). If a particular channel \( c \) in the dataset \( \mathbf{X}_c \) contains missing entries, the MICE approach is applied to impute the data.
Mathematically, the imputation process models each feature column \( \mathbf{x}_j \) conditionally on the others:
\begin{equation}\label{eq:pij}
P(\mathbf{x}_j \mid \mathbf{x}_{-j}) \propto \prod_{i=1}^N P(x_{ij} \mid \mathbf{x}_{i, -j}),
\end{equation}
where \( \mathbf{x}_{-j} \) denotes all columns except \( j \), and \( x_{ij} \) is the value at row \( i \), column \( j \).  
The MICE algorithm estimates missing entries iteratively. At each step \( t \), the missing value \( x_{ij} \in \mathbf{\kappa} \) is estimated as Equation~\eqref{eq:fij}:
\begin{equation}\label{eq:fij}
x_{ij}^{(t)} = f(X_{-j} | \theta_j^{(t)}) + \epsilon_{ij}^{(t)},
\end{equation}
where \( f \) is a predictive model trained on the observed data \( X_{-j} \) (all features excluding \( j \)), \( \theta_j^{(t)} \) are the parameters of the model at iteration \( t \), and \( \epsilon_{ij}^{(t)} \sim \mathcal{N}(0, \sigma^2) \) adds randomness to account for uncertainty. 
The algorithm iterates until convergence, producing a complete dataset \( \hat{X} \) or \( \mathbf{X}_{c, \text{imputed}} \), where imputed values stabilize. This ensures a complete dataset ready for subsequent feature extraction steps.

\subsubsection{Singular Value Decomposition (SVD)}

SVD decomposes \( \mathbf{X}_{c, \text{imputed}} \) as given in Equation~\eqref{eq:imp}:
\begin{equation}\label{eq:imp}
\mathbf{X}_{c, \text{imputed}} = \mathbf{U} \mathbf{S} \mathbf{V}^\top,
\end{equation}
where \( \mathbf{U} \in \mathbb{R}^{N \times N} \): Left singular vectors, \( \mathbf{S} \in \mathbb{R}^{N \times M} \): Diagonal matrix of singular values \( \sigma_i \), \( \mathbf{V}^\top \in \mathbb{R}^{M \times M} \): Transposed right singular vectors.
The singular values \( \sigma_i \) are ordered \( \sigma_1 \geq \sigma_2 \geq \cdots \geq \sigma_{\min(N, M)} \).
Although the singular values are monotonically decreasing, their magnitudes do not correspond linearly to the amount of information captured. 
Instead, the variance (or energy) associated with each component is proportional to $\sigma_i^2$. 
Consequently, a singular value twice as large as another reflects roughly four times the variance contribution, rather than twice the information. 
This interpretation underpins the use of cumulative variance ratios when selecting the optimal number of components in our dimensionality reduction step.

 Selecting the top \( r \) components retains the most significant features to obtain Equation~\eqref{eq:top}:
\begin{equation}\label{eq:top}
\mathbf{X}_{c, r} = \mathbf{U}_r \mathbf{S}_r \mathbf{V}_r^\top,
\end{equation}
where \( \mathbf{U}_r \in \mathbb{R}^{N \times r} \),
\( \mathbf{S}_r \in \mathbb{R}^{r \times r} \),
\( \mathbf{V}_r^\top \in \mathbb{R}^{r \times M} \).

To reconstruct \( \mathbf{X}_{c, \text{imputed}} \) from the top components we have Equation~\eqref{eq:recon}:
\begin{equation}\label{eq:recon}
\mathbf{\hat{X}}_{c} = \mathbf{U}_r \mathbf{S}_r \mathbf{V}_r^\top.
\end{equation}
This reconstruction minimises the error while retaining variance, defined by Equation~\eqref{eq:evr}:
\begin{equation}\label{eq:evr}
\text{Explained Variance Ratio} = \frac{\sum_{i=1}^r \sigma_i^2}{\sum_{i=1}^{\min(N, M)} \sigma_i^2}.
\end{equation}

\noindent The coefficient matrix is expressed as Equation~\eqref{eq:coef}:
\begin{equation}\label{eq:coef}
\mathbf{C} = \mathbf{U}_r \mathbf{S}_r,
\end{equation}
serves as the feature set for classification. These features encapsulate the geometric and intensity-based properties of radio galaxy morphologies, enabling differentiation between FR-I, FR-II, Compact, and Bent types.

\begin{figure}
\begin{minipage}[H]{\linewidth}
\centering
\includegraphics[width=\textwidth]{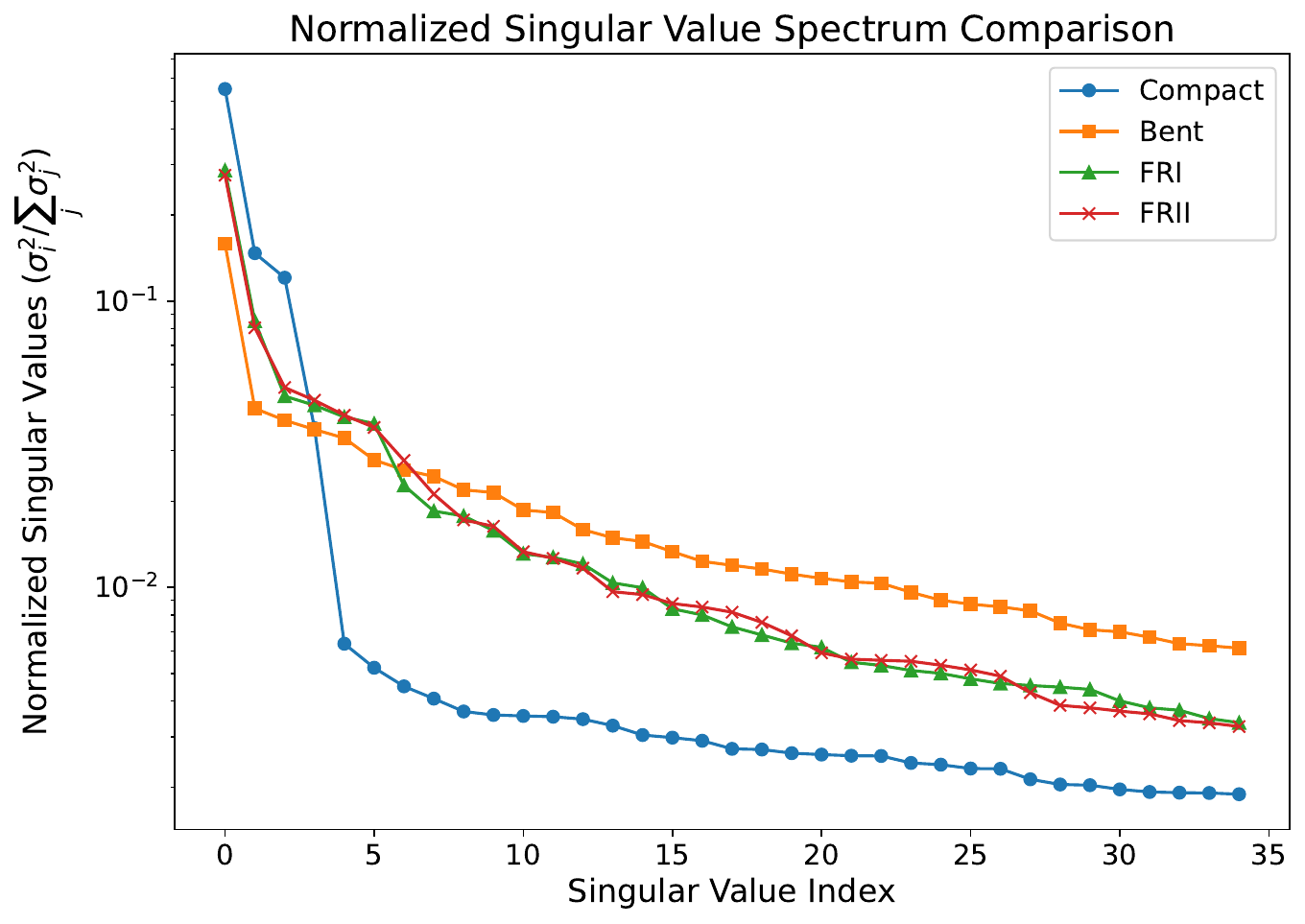} 
\end{minipage}
\caption{Singular value decay profiles for four morphological types of radio galaxies: Compact (blue circles), Bent (orange squares), FRI (green triangles), and FRII (red crosses). The y-axis represents the normalised singular values, and the x-axis corresponds to the rank of the singular value components. Compact galaxies exhibit a sharp decline, indicative of their simple morphology. Bent galaxies show a slower decay, reflecting their intermediate complexity due to jet bending. FRI galaxies exhibit a moderate decay rate, representing their diffuse and gradual jet structures, while FRII galaxies display the slowest decay, highlighting their intrinsic structural complexity due to prominent lobes and hotspots.
\textbf{Note:} Singular values are shown in normalised form ($\sigma_i^2 / \sum_j \sigma_j^2$) to ensure comparability across subsets, since raw singular values are scale-dependent.
}\label{fig:_svdplot_}
\end{figure}

\begin{figure*}
\begin{minipage}[H]{\linewidth}
\centering
\includegraphics[width=\textwidth]{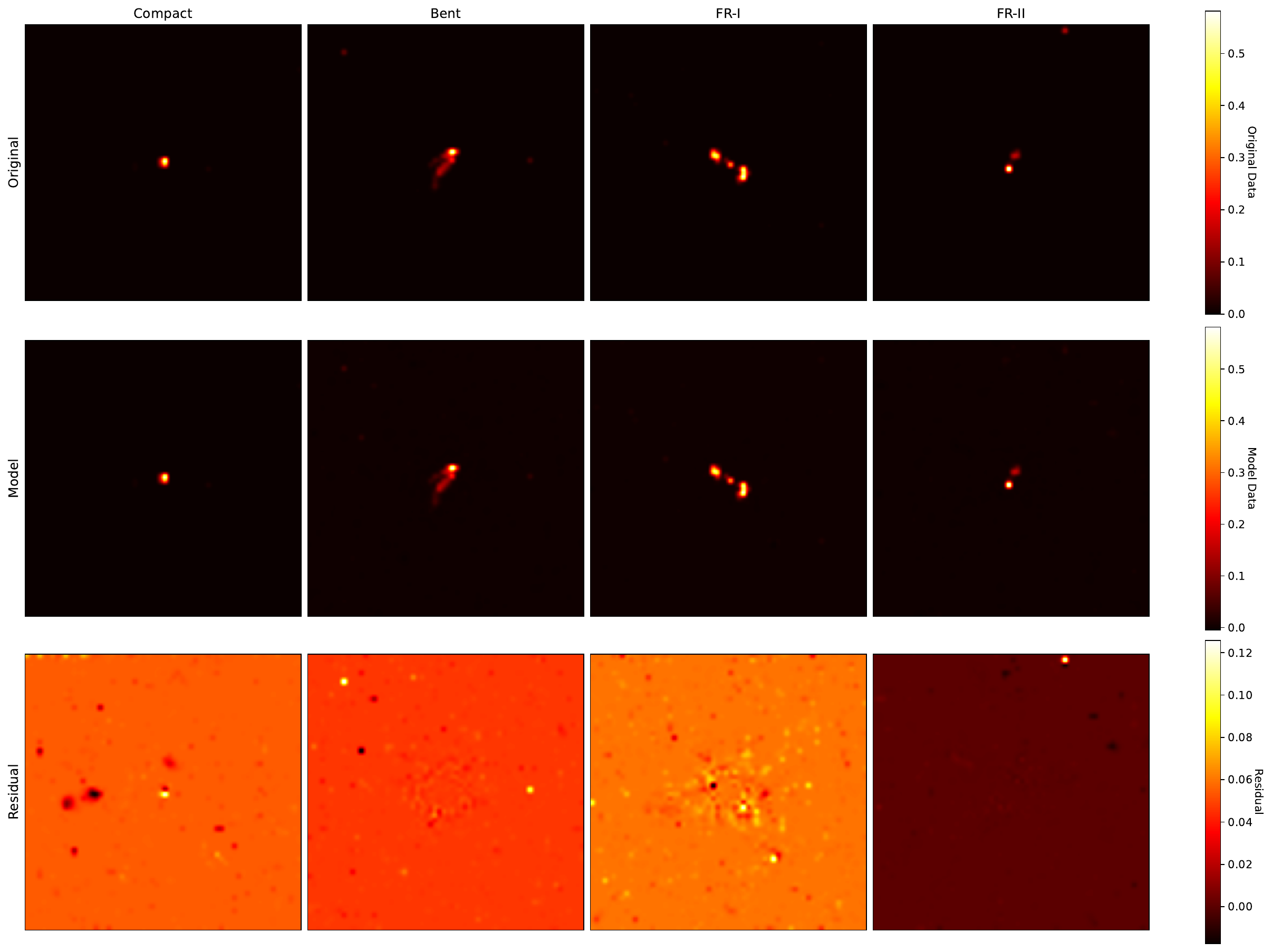} 
\end{minipage}
\caption{Reconstruction of radio galaxy images using the top $120$ singular values out of 16,384. The first two rows correspond to reconstructed images for various galaxy types (Compact, Bent, FRI, and FRII), showcasing the retention of core morphological features such as jets, lobes, and hotspots. The bottom row displays the residual differences between the original and reconstructed images, highlighting the absence of finer details and noise in the reconstructions. The colour bar indicates the normalised intensity levels, with brighter regions representing higher energy emissions. This reconstruction emphasises the dominant structures while effectively suppressing noise and minor variations, illustrating the utility of Singular Value Decomposition for dimensionality reduction in astronomical imaging.
\textbf{Note:} The reconstructed, residual, and original image stamps are $128 \times 128$ pixels, corresponding to an angular extent of roughly $230^{\prime\prime} \times 230^{\prime\prime}$ (i.e., $\approx 3.8^{\prime} \times 3.8^{\prime}$), using the FIRST survey pixel scale of $1.8^{\prime\prime}$ per pixel.
}\label{fig:svd_recon}
\end{figure*}

The singular value profiles depicted in Fig.\ref{fig:_svdplot_} reveal the complexity and variability inherent in the morphologies of the four types of radio galaxies: Compact, Bent, FRI, and FRII.
The singular value curve for Compact radio galaxies shows a sharp decline, with a rapid saturation of the singular values. This indicates that the majority of the variance in the data is captured by the first few singular components, which is consistent with the relatively simple and symmetric nature of these sources. 
It is important to note that singular values are data-dependent and not directly comparable across unnormalised subsets. 
Their magnitudes depend on the overall variance and scaling of each subset. 
Therefore, in Figure~\ref{fig:_svdplot_}, we present normalised singular values, expressed as $\sigma_i^2 / \sum_j \sigma_j^2$, which reflect the relative variance contribution of each component and allow meaningful comparison across different data partitions.

Compact radio galaxies exhibit minimal extended structure, and their emission is confined to small angular scales, making them highly amenable to low-rank approximations.
Bent radio galaxies show a singular value profile with a slower decline compared to Compact sources. The bending of jets, often due to interactions with the surrounding medium, introduces structural complexity, reflected in the higher contribution of additional singular components. While the dominant modes still account for a significant fraction of the variance, the tail of the singular value distribution remains elevated, indicating the presence of intermediate-scale features that require additional components to describe accurately
The profile for FRI radio galaxies exhibits a moderate decay rate of singular values, reflecting their characteristic diffuse and large-scale emission. FRIs are typically associated with jets that fade with increasing distance from the core, producing gradual transitions in intensity. This structural complexity is encoded in higher singular values compared to Compact and Bent sources. The slower decay suggests that capturing the full morphological diversity of FRI galaxies necessitates a broader set of basis vectors.
The singular value curve for FRII galaxies displays the slowest decay among the four classes, highlighting their intrinsic complexity. The prominent lobes and bright hotspots of FRIIs introduce significant variability on large angular scales. This necessitates the inclusion of numerous singular components to effectively represent their morphological features. The extended tail of the singular value distribution suggests that these galaxies are rich in structural details, which are spread across both large and small scales.

The reconstruction of radio galaxy images using SVD is illustrated in Fig.~\ref{fig:svd_recon}. This technique enables dimensionality reduction while retaining critical structural information.
The original images (Row $1$) display intricate details and intensity variations characteristic of different radio galaxy types. In contrast, the reconstructed images (Row $2$), generated using the top $120$ singular values, retain dominant morphological features like galaxy cores, jets, and lobes.
The residual differences (Row $3$) highlight the fine details and noise lost during truncation. The top 120 singular values capture the majority of variance in the original data, prioritising dominant spatial features while omitting high-frequency noise and smaller emission variations.
This approach achieves a dramatic reduction in dimensionality, using less than $1\%$ of the available singular values, making it computationally efficient and effective for large-scale radio galaxy datasets.

\subsection{Addressing Class Imbalance} \label{sec:class_bal}

Radio galaxy classification presents a significant challenge due to the inherent class imbalance across galaxy types. Our dataset includes four classes, each with varying sample sizes.
This imbalance can lead to model bias, where the classifier favours the majority class. 
To mitigate this, we implemented a LNE\footnote{Local Neighbourhood Encodings} pipeline--a hybrid method inspired by the work of \cite{koziarski2024local} that aims to generate a balanced dataset while preserving the astrophysical characteristics of each class.

The core of the LNE pipeline involves adaptive resampling the dataset to mitigate class imbalance. Let the dataset comprise \( C \) classes, where each class \( c \) has \( N_c \) samples. Unlike traditional approaches that fix a uniform target class size \( T \), LNE dynamically determines resampling intensities for each class as part of an optimisation process. Specifically, LNE encodes the strength of oversampling and undersampling through a learnable vector, which is optimised using evolutionary strategies with respect to a chosen performance metric (e.g., F1-score, balanced accuracy).
This allows the final class distribution to emerge from the data itself rather than enforcing arbitrary balance, improving both flexibility and classification fidelity.
As an optional preprocessing step, One-Class SVM is employed to detect and remove potential outliers in the dataset before the balancing process. The algorithm identifies samples that deviate significantly from the feature distribution of the majority of samples in a class. This ensures that the resampling methods are applied to a refined dataset with fewer noisy or irrelevant samples, maintaining the astrophysical validity of the radio galaxy classification features.  
For each class \( c \), One-Class SVM operates on the subset \( X_c \) as follows:  
\begin{equation}\label{eq:onesvm}
X_c' = \{\mathbf{x} \in X_c : f(\mathbf{x;\epsilon}) = +1\},
\end{equation}
where \(f(\mathbf{x;\epsilon})\) denotes the decision function of the One-Class SVM with contamination parameter \( \epsilon \).
The parameter $\epsilon \in [0, 1)$ explicitly controls the expected fraction of observations treated as outliers during model fitting;
smaller values retain more samples, whereas larger values remove a higher proportion of potential anomalies.
Thus, contamination directly modulates the refinement of the class subset \( X_c' \) before resampling.

Depending on the initial size of class \( c \), two resampling methods are applied: oversampling for minority classes and undersampling for majority classes.
For classes where \( N_c < T \), new synthetic samples are generated using a k-nearest neighbours (KNN) approach.
For each sample \( \mathbf{x} \in X_c \), a neighbor \( \mathbf{x}' \) is randomly selected, and a synthetic sample \( \tilde{\mathbf{x}} \) is generated by:
\begin{equation}\label{eq:lne_0}
\tilde{\mathbf{x}} = \mathbf{x} + \eta, \quad \eta \sim \mathcal{N}(0, \sigma^2),
\end{equation}
where \( \sigma = 0.1 \cdot \|\mathbf{x} - \mathbf{x}'\|_2 \). 
This ensures that the synthetic sample \( \tilde{\mathbf{x}} \) lies within the local manifold of the feature space of class \( c \), maintaining the inherent structure of the class. The addition of noise \( \eta \sim \mathcal{N}(0, \sigma^2) \) ensures that the generated samples are sufficiently distinct while preserving the class's feature distribution.
For classes where \( N_c > T \), undersampling is applied to reduce the class size. This is achieved by randomly selecting a subset of the original class samples:
\begin{equation}\label{eq:lne_1}
X_c' = \text{Resample}(X_c, \text{n\_samples}=T, \text{replace=False}).
\end{equation}
This ensures that the resulting balanced dataset contains \( T \) samples from each class, with no duplication for majority classes. 
The final dataset, denoted \( X' \), is obtained by combining all resampled classes, such that each class \( c \) has \( T \) samples.

\begin{algorithm}
\caption{Local Neighbourhood Encodings for Class Balancing}\label{alg:lne}
\begin{algorithmic}[1]
\Require Dataset \( X \), labels \( y \), fitness metric \( \mathcal{F} \), generations \( G \), population size \( P \), contamination rate \( \epsilon \)
\Ensure Resampled dataset \( X', y' \)

\State \textbf{Step 1: Preprocessing}
\State \quad Impute missing values in \texttt{X} using an imputation method (e.g., MICE)
\For{each class \( c \) in \( y \)}
    \State Apply One-Class SVM to filter outliers from \( X_c \) with contamination \( \epsilon \) \label{step:iv}
    \State \( X_c' = \{\mathbf{x} \in X_c : f(\mathbf{x;\epsilon}) = +1\} \)
\EndFor
\State Merge all \( X_c' \) to form filtered dataset \( X_{\text{filtered}} \)
\State Identify majority and minority classes in \( y_{\text{filtered}} \)
\State Compute KNN for all samples in \( X_{\text{filtered}} \)
\State Classify each sample as Safe, Borderline, Rare, or Outlier (neighbourhood-based)

\State \textbf{Step 2: Initialise Population}
\State Initialise resampling strategy vectors \( E \in \{0, 1, 2\}^n \) across \( P \) individuals
\State Encoding: 0 = retain, 1 = oversample, 2 = remove

\For{generation \( g = 1 \) to \( G \)}
    \For{each individual \( E \)}
        \State Apply resampling strategy to \( X_{\text{filtered}}, y \)
        \State Use KNN to generate synthetic samples for oversampled instances
        \State Evaluate classifier on resampled data; compute fitness \( \mathcal{F}(E) \)
    \EndFor
    \State Select top-performing individuals; apply mutation and crossover
\EndFor

\State \textbf{Step 3: Final Dataset}
\State Apply best resampling strategy \( E^* \) to obtain \( X', y' \)
\State \Return \( X', y' \)
\end{algorithmic}
\end{algorithm}

Traditional methods for addressing class imbalance, such as SMOTE\footnote{Synthetic Minority Over-sampling Technique} \citep{chawla2002smote}, Cluster Centroids \citep{leisch2006toolbox}, and Tomek Links (available in the {\tt imbalanced-learn}\footnote{https://imbalanced-learn.org/stable/} library) \citep{1976ITSMC...6..121T}, have some limitations when applied to radio galaxy classification. 
For instance, SMOTE generates synthetic samples by interpolating between nearest neighbours, which can introduce unrealistic samples that may not reflect the underlying astrophysical processes.
Given the nonlinear nature of radio galaxy features, this approach risks introducing artefacts that distort class distributions \citep{schawinski2017generative,dieleman2015rotation}.
Also, Cluster Centroids simplifies the class distributions by reducing the number of samples through centroids of clustered data points. 
While this method can reduce imbalance, it may oversimplify complex feature patterns, leading to a loss of important morphological details, such as those in the radio galaxy classification task.
Tomek Links on the other hand, focuses on removing boundary-level samples to clean the decision boundaries. Still, it fails to address severe class imbalance and may ignore spatial correlations inherent in astrophysical datasets.
In contrast, the LNE method preserves the local structure of the feature space, which is crucial to maintaining the intrinsic astrophysical characteristics of each class. By synthesising samples within local neighborhoods, LNE ensures that the generated data points are both realistic and meaningful, avoiding the artifacts that can arise from other methods like SMOTE. Furthermore, LNE adapts dynamically to the size of each class, ensuring a balanced dataset without overcompensating for the imbalance.
The pseudocode in Algorithm~\ref{alg:lne} outlines the key steps of the LNE method, including our custom integration of One-Class SVM for outlier filtering. 
The contamination parameter $\epsilon$ directly influences Step~\ref{step:iv} of Algorithm \ref{alg:lne} by defining the percentage of samples removed as outliers prior to resampling, thereby controlling contamination in the class-balancing procedure.

\begin{figure*}
\begin{minipage}[H]{\linewidth}
\centering
\includegraphics[width=\textwidth]{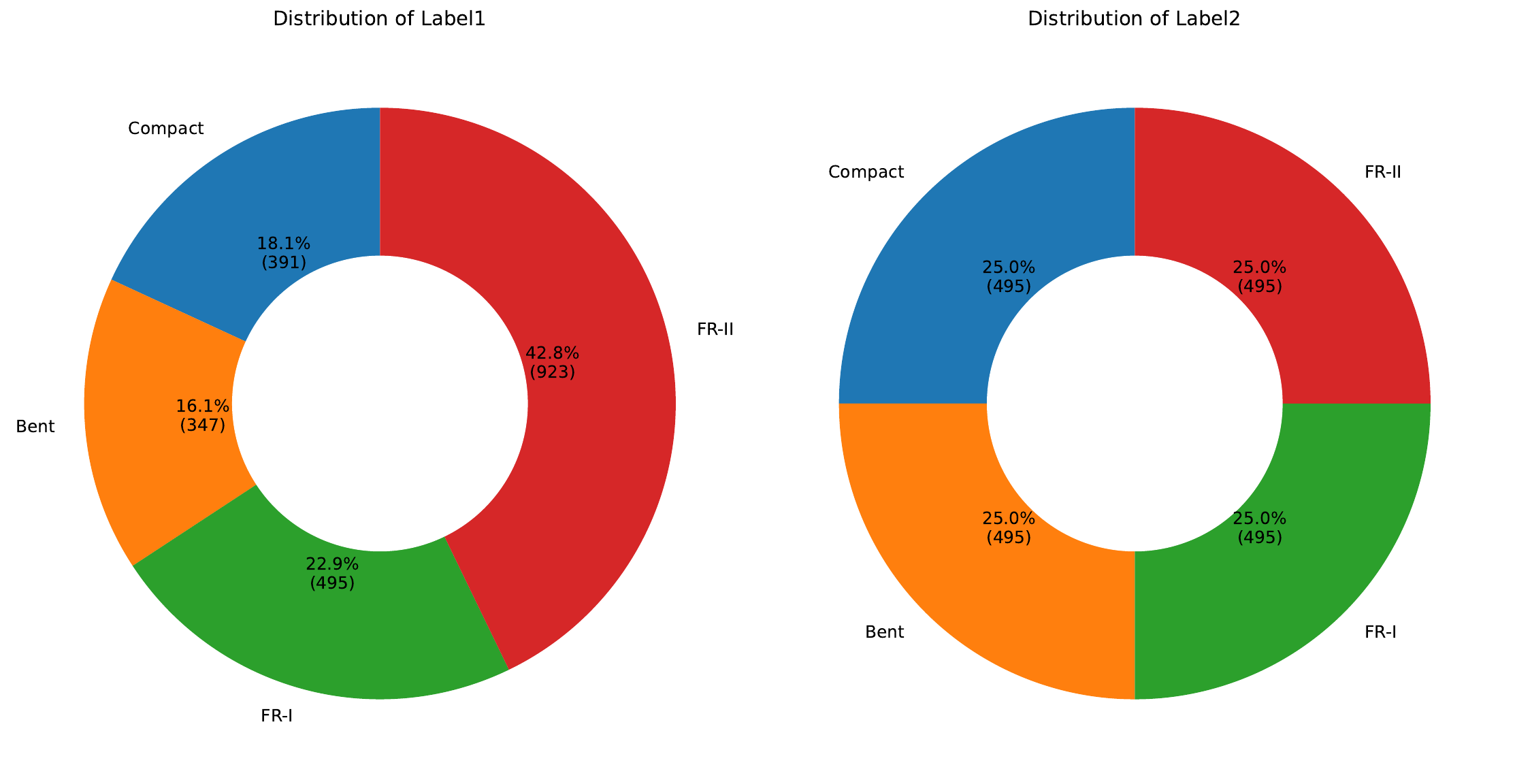} 
\end{minipage}
\caption{Class distribution of radio galaxy categories before and after applying the LNE algorithm. The initial distribution demonstrates class imbalance (refer to the left plot), with varying sample sizes across classes. After balancing, the distribution is uniform (refer to right plot), with each class containing an equal number of samples, facilitating unbiased model training and improved classification performance.}\label{fig:_LocalNeighborhoodEncodings_}
\end{figure*}

Fig.~\ref{fig:_LocalNeighborhoodEncodings_} illustrates the class distribution of the radio galaxy dataset before and after applying the LNE algorithm.
The imbalance displayed in the left pie chart might lead to the calssification model being better at predicting the more prevalent class (FR-II) and less effective at predicting the less represented classes (Compact and Bent).
After applying the LNE algorithm (right pie chart), the class distribution is adjusted to achieve balance across all classes.
Here, the LNE algorithm has resampled the dataset by either over-sampling the minority classes (Compact and Bent) or under-sampling the majority class (FR-II), so that each class now has an equal number of $495$ samples.
This balance is essential for training machine learning models as it ensures the model is exposed to an equal representation of all classes, which helps to prevent bias towards the more prevalent classes. 

\subsection{Dataset Partitioning and Validation Strategy}

To evaluate the performance of our classification framework and prevent overfitting or data leakage, we adopted a rigorous dataset splitting strategy. After applying SVD for feature extraction, the decomposed dataset for all classes consisted of an original shape of $(2158, 120)$. To address class imbalance bias, we employed the LNE technique, which resulted in a balanced dataset of shape $(1980, 120)$.
From this balanced dataset, we randomly shuffled the instances to eliminate any order bias. Subsequently, we partitioned the data into training and testing sets using an $80/20 $ split--80\% of the samples (approximately 1584 instances) were designated for training, while the remaining 396 instances were reserved for testing. This splitting was performed after shuffling to ensure randomness and was stratified across classes to preserve proportional class distributions in both sets.

All data preprocessing, including the shuffling, splitting, and feature extraction steps, was conducted only on the training dataset within each cross-validation fold to prevent data leakage. The model was trained solely on the training data, and its performance was evaluated independently on the unseen test data. This approach ensures that the classification results are robust and generalizable, effectively mitigating the risk of overfitting.

\subsection{Baseline Models}

The baseline classifiers employed in this study--Logistic Regression (LogisticRegr), Support Vector Machine (SVM), Random Forest (RF), LightGBM, and Multilayer Perceptron (MLP) were selected for their complementary inductive biases and proven performance in high-dimensional classification tasks, particularly under class imbalance and morphological variation. 
These models form the core constituents of the Bayesian ensemble framework and were chosen to provide a balanced trade-off between interpretability, robustness, and representational capacity. To ensure consistency across learners, all hyperparameters were jointly optimised using Particle Swarm Optimisation (PSO) \citep{eberhart1995particle}.

LogisticRegr was included for its stability, low variance, and interoperability. 
Certain radio galaxy subclasses, particularly Compact sources, exhibit quasi-linear separability in SVD-reduced space, making the logistic model a suitable baseline classifier. Its calibrated probabilistic outputs also support downstream weighting in the Bayesian ensemble. The regularisation parameter \(C\) was tuned within the PSO framework.

SVM complements LogisticRegr by offering capacity for non-linear class separation. We employed an RBF-kernel SVM to capture complex boundaries, particularly in cases where FR-I and Bent morphologies overlap. The choice reflects SVM’s robustness in sparse, high-dimensional data, as well as prior use in astronomical classification tasks. The parameters \(C\) and \(\gamma\) were optimised through the unified PSO routine.

LightGBM was selected as a base learner for its efficiency, robustness to class imbalance, and ability to model complex, non-linear interactions in sparse, high-dimensional feature spaces such as those derived from truncated SVD. Its leaf-wise tree growth strategy facilitates finer-grained partitioning compared to level-wise boosting methods, which is advantageous in capturing subtle morphological variations. Key hyperparameters, including learning rate and number of leaves, were jointly optimised using PSO.
Importantly, LightGBM’s native support for SHAP \citep{lundberg2017unified,shapley1953value} values enabled post hoc analysis of feature contributions, providing insights into which SVD components were most influential in class differentiation.

MLP was included to capture non-linear and higher-order feature interactions that may be inaccessible to tree-based methods. A compact architecture with ReLU activation and dropout was selected to mitigate overfitting, particularly given class imbalance in the training data. Hyperparameters such as hidden layer size, solver, activation function, and learning rate schedule were tuned using PSO. 

Finally, in contrast to the other base learners, RF was employed solely as the meta-learner in the stacking ensemble. Its use was motivated by its robustness to overfitting, capacity to aggregate heterogeneous predictive outputs, and empirical success in similar hierarchical model architectures. As the final estimator, RF integrated the level-one predictions from all optimised base learners, with its hyperparameters tuned within the same PSO framework to ensure compatibility and performance stability across cross-validation folds.

Taken together, these five classifiers provide a deliberate balance of inductive biases and computational properties. Their inclusion reflects an emphasis on ensemble diversity, model complementarity, and uncertainty calibration. Further algorithmic details are available in the cited literature; our focus here is on each model’s integration and role within the proposed classification framework.

\subsection{Particle Swarm Optimisation for Hyperparameter Tuning} \label{sec:psohyper}

PSO is an effective optimisation method inspired by the collective behaviour of birds and fish. 
It is particularly adept at hyperparameter tuning because it can explore complex search spaces without relying on gradient information.
Within this work, PSO is used to optimise the hyperparameters for four base models: LogisticRegr, SVM, LightGBM, and MLP.
The PSO algorithm is initialised by creating a population of candidate solutions known as particles, which are randomly positioned across the search space. 
Each particle symbolises a possible configuration of hyperparameters and possesses a position vector \( \vec{x_i} \) and a velocity vector \( \vec{v_i} \).
The positions of the particles are updated iteratively, influenced by their individual best-known position \( \vec{p_i} \), as well as the global best-known position \( \vec{g} \) found by any particle in the swarm. The formula for updating velocity is given by Equation~\eqref{eq:pso_0}:
\begin{equation}\label{eq:pso_0}
\vec{v_i}(t+1) = w \vec{v_i}(t) + c_1 r_1 (\vec{p_i} - \vec{x_i}) + c_2 r_2 (\vec{g} - \vec{x_i}),
\end{equation}
where \( w \) is the inertia weight controlling the impact of the previous velocity, \( c_1 \) and \( c_2 \) are acceleration coefficients, and \( r_1 \) and \( r_2 \) are random numbers sampled uniformly from \([0, 1]\). The position is then updated as Equation~\eqref{eq:pso_1}:
\begin{equation}\label{eq:pso_1}
\vec{x_i}(t+1) = \vec{x_i}(t) + \vec{v_i}(t+1).
\end{equation}

\noindent This iterative process balances exploration (searching new areas of the space) and exploitation (refining existing good solutions), enabling PSO to converge towards optimal hyperparameters.
This is governed by three key parameters: the inertia weight (\( w \)), personal acceleration coefficient (\( c_1 \)), and social acceleration coefficient (\( c_2 \)). The inertia weight (\( w \)) controls how much of the previous velocity of the particle influences its current movement. In this study, \( w \) decreased linearly from $0.8$ to $0.4$ over iterations, promoting exploration in the early stages and exploitation as convergence approached. 
The coefficients \( c_1 \) and \( c_2 \), both set to $1.5$, guided particles towards their personal best positions and the global best position, balancing individual learning and social collaboration. A swarm size of \( 20 \) particles and \( 50 \) iterations was selected based on a combination of insights from prior literature and empirical testing. Previous studies applying PSO to similar optimisation tasks in high-dimensional search spaces have shown that swarm sizes in the range of $20$--$40$ often offer a good trade-off between exploration and convergence speed \citep{clerc2010particle,eberhart1995particle}. To verify suitability in our specific context, we performed preliminary tuning experiments with swarm sizes of $10, 20,$ and $40$, and iteration limits ranging from $30$ to $100$. The chosen configuration (\(20\) particles, \(50\) iterations) demonstrated consistent convergence across folds without incurring excessive computational cost. This setting offered stable optimisation performance while avoiding premature convergence, a risk associated with smaller swarms and fewer iterations. 

The cost function for each model was the average cross-validation accuracy, ensuring that the optimised hyperparameters generalised well to unseen data. This careful design and tuning of the PSO parameters minimised the risk of premature convergence or excessive exploration, allowing the algorithm to efficiently locate optimal solutions in the defined search spaces.

For Logistic Regression, the single parameter \( C \), which controls regularisation, is optimised. The cost function evaluates the average cross-validation accuracy for a given \( C \), and PSO searches the space \([0.01, 10.0]\) to identify the value that maximises accuracy. 
For SVM, the optimisation involves two parameters: \( C \), which controls the regularisation strength, and \( \gamma \), the kernel coefficient for the radial basis function (RBF) kernel. The search space for \( C \) and \( \gamma \) is defined as \([0.01, 10.0]\) and \([0.001, 1.0]\), respectively. Each particle in the swarm evaluates a pair of \( (C, \gamma) \), and the cost function computes the average accuracy from cross-validation. 
LightGBM tuning is more complex, involving two parameters: the number of leaves \( \text{num\_leaves} \) and the learning rate \( \eta \). The search space for \( \text{num\_leaves} \) is \([10, 100]\), and for \( \eta \), it is \([0.01, 0.5]\). PSO navigates this multi-dimensional space, using the cross-validation accuracy as the cost function to identify the optimal combination of parameters.
For the Multi-Layer Perceptron (MLP), a flexible and highly parametric model, PSO was used to optimise both continuous and categorical hyperparameters. The hyperparameters included the size of the hidden layers, activation function, solver type, learning rate schedule, and the maximum number of iterations. The hidden layer configurations were chosen from four predefined options: \((50,)\), \((100,)\), \((50, 50)\), and \((100, 50)\), reflecting a mix of single- and multi-layer architectures. Activation functions were restricted to \(relu\) and \(tanh\), ensuring compatibility with the model's inherent capabilities. For the solver, three choices were provided: \(lbfgs\), \(sgd\), and \(adam\), offering a balance between optimisation speed and adaptability. The learning rate schedule options included \(constant\), \(invscaling\), and \(adaptive\), allowing the model to adapt to varying convergence requirements.
The optimisation process utilised a cost function that evaluated the cross-validation accuracy of each hyperparameter configuration. PSO encoded the categorical variables (e.g., activation function, solver type, and learning rate schedule) as indices within their respective ranges, while the maximum number of iterations was treated as a continuous variable. Each particle in the swarm represented a combination of these parameters, and the algorithm iteratively refined the configurations to maximise model performance. For example, a particle’s vector might represent the indices for a hidden layer configuration of \((50, 50)\), the activation function \(relu\), the solver \(adam\), the learning rate schedule \(adaptive\), and a maximum iteration count of $200$. By systematically exploring this diverse search space, PSO was able to identify configurations that achieved optimal cross-validation accuracy for the MLP.

\subsection{Traditional Ensemble Learning} \label{sec:TEL}

Ensemble learning methods, such as Averaging, Boosting, Bagging, and Stacking, represent distinct but related strategies to combine the predictions of multiple base models to improve classification performance.
These methods are widely used in machine learning tasks due to their ability to reduce variance, bias, and enhance generalisation performance.
In the following, we discuss the mathematical foundations and principles behind these methods.

\subsubsection{Averaging Ensemble} \label{sec:TEL_avg}

The Averaging ensemble method is a simple yet powerful approach that calculates the mean of the predicted probabilities of multiple base models and derives the final class prediction from the maximum averaged probability. Mathematically, for \( N \) base models \( f_i(x) \), the ensemble prediction is given by Equation~\eqref{eq:aven}:

\begin{equation}\label{eq:aven}
P(y = c | x) = \frac{1}{N} \sum_{i=1}^N P_i(y = c | x),
\end{equation}

\noindent  where \( P_i(y = c | x) \) represents the probability that the \( i \)-th model predicts class \( c \). Averaging tends to work well when the base models are independently trained and exhibit diverse predictions, as it minimises the variance of the ensemble by balancing the individual model predictions. 
Research such as \cite{opitz1999popular} emphasizes that diversity among models is crucial for averaging to be successful, since very similar models result in minimal gains compared to using an individual classifier.

\subsubsection{Boosting Ensemble} \label{sec:TEL_boost}

Boosting is an ensemble method that incrementally adjusts the weights of base models according to how well they perform. The main aim is to focus more on the instances that were misclassified during each iteration, allowing future models to improve upon the weaknesses of the previous ones. This step-by-step correction system renders boosting an effective strategy for enhancing predictive accuracy in intricate tasks.
The mathematical foundation of boosting is characterised by a weighted combination of base model predictions, with the weights being dynamically altered to represent model effectiveness. For \( N \) base models, the ensemble prediction is represented as Equation~\eqref{eq:boost_norm_0}:

\begin{equation}\label{eq:boost_norm_0}
P(y = c | x) = \sum_{i=1}^N \alpha_i P_i(y = c | x),
\end{equation}

\noindent   where \( \alpha_i \) denotes the weight assigned to the \( i \)-th model, and \( P_i(y = c | x) \) represents the probabilities predicted for class \( c \) by the \( i \)-th model. In traditional methods such as AdaBoost \citep{freund1997decision}, these weights are explicitly updated based on the model's classification error rate \( \epsilon_i \) as defined in Equation~\eqref{eq:boost_norm_1}:

\begin{equation}\label{eq:boost_norm_1}
\alpha_i = \log \frac{1 - \epsilon_i}{\epsilon_i}.
\end{equation}

\noindent   This formula ensures that models with lower error rates are given higher weights, emphasizing their contributions to the final prediction. AdaBoost also formalised the theoretical guarantees of boosting, demonstrating its ability to reduce both training error and overfitting through a careful trade-off between model complexity and empirical risk minimisation.
In this work, the boosting process was simplified by directly scaling weights based on performance metrics, such as accuracy.
This practical adaptation avoids the complexity of explicitly computing error rates while preserving the core principle of emphasising models that perform better. During each iteration, the weighted predictions of all base models are aggregated, and the ensemble's output probabilities are normalized to ensure stability as presented in Equation~\eqref{eq:boost_norm_2}:

\begin{equation}\label{eq:boost_norm_2}
\hat{P}(y = c | x) = \frac{\sum_{i=1}^N \alpha_i P_i(y = c | x)}{\sum_{i=1}^N \alpha_i}.
\end{equation}

\noindent This normalisation ensures that the aggregated probabilities sum to $1$ across all classes. The final prediction is obtained by selecting the class with the highest normalised probability.

\subsubsection{Bagging Ensemble} \label{sec:TEL_bag}

Bagging, or bootstrap aggregating, reduces model variance by training each base model on a different bootstrap sample of the training data and averaging their predictions as the mathematical formulation in Equation~\eqref{eq:aven}.
Bootstrap sampling ensures that each base model is trained on a slightly different subset of the data, introducing diversity into the ensemble. This diversity helps to reduce overfitting, especially in high-variance models such as decision trees.
In Bagging, increasing the number of models in the ensemble leads to a reduction in variance, as explained by the law of large numbers.

\subsubsection{Stacking Ensemble} \label{sec:TEL_stack}

Stacking combines the strengths of multiple models by training a meta-model to aggregate their predictions. In the first step, the base models are trained to generate predictions or probability distributions. These predictions are then used as input features for a higher-level meta-model, which learns how to optimally combine the base models. Mathematically, if \( f_i(x) \) represents the output of the \( i \)-th base model, the meta-model learns the function \( g \) as expressed in Equation~\eqref{eq:stack_en}:

\begin{equation}\label{eq:stack_en}
y = g(f_1(x), f_2(x), \ldots, f_N(x)).
\end{equation}

\noindent  The final prediction is derived from the meta-model \( g \), which in our case we adopted RF algorithm. Stacking has proven particularly effective in leveraging the complementary strengths of diverse models, as shown by \cite{wolpert1992stacked} in his foundational work on stacked generalisation. The method is highly suitable for radio galaxy classification tasks, where complex decision boundaries and multi-class predictions benefit from a hierarchical combination of models.

\subsection{Bayesian Ensemble Learning} \label{sec:BEL}

Bayesian ensemble learning represents an advanced methodological shift in how ensemble models are conceptualised and applied, particularly in multiclass classification tasks.
Traditional ensemble approaches, such as Averaging, Boosting, Bagging, and Stacking, often assume homogeneity in the contribution of baseline models to the ensemble’s predictive output.
This is reflected in the uniform weighting or static aggregation rules frequently employed, where all models are treated as equally reliable contributors. However, this assumption becomes problematic when the baseline models exhibit significant differences in performance.
In real-world scenarios, including automated radio galaxy classification, certain models are inherently more effective at capturing specific patterns or class distinctions because of their design or learning paradigms.
Bayesian ensemble learning addresses these challenges by dynamically incorporating model-specific performance into the weighting process, ensuring that the ensemble reflects the relative reliability of each baseline model.
The critical justification for adopting Bayesian ensemble learning lies in its ability to assign probabilistic weights that evolve based on empirical evidence. 
Unlike traditional methods, Bayesian ensemble learning integrates Gamma priors, Bayesian updating, and weight regularisation to compute model weights. 
These elements collectively ensure that the ensemble captures heterogeneity in model performance and adapts dynamically to variations between classes. This approach aligns with findings from recent literature emphasising the importance of incorporating model uncertainty and performance metrics into ensemble frameworks.
For instance, studies by \citet{kapoor2023probabilistic,ghahramani2015probabilistic}  highlight how Bayesian inference provides a principled framework for uncertainty quantification in machine learning, enabling better decision-making in complex tasks. 

The use of probabilistic weights, derived from Gamma priors, offers several advantages. First, the flexibility of the Gamma distribution allows prior beliefs about model reliability to be encoded explicitly, reflecting domain knowledge or empirical observations. Unlike Uniform priors, which assume equal reliability across all models, Gamma priors accommodate heterogeneity, assigning higher initial weights to models expected to perform better. This is particularly valuable in multiclass classification, where certain models may excel in distinguishing specific classes while others provide complementary strengths. For example, in radio galaxy classification, a LightGBM model may excel at identifying Compact galaxies due to its ability to capture subtle feature interactions, while an SVM may be better suited for distinguishing between FR-I and FR-II galaxies with clear boundary separations.
The weight \( w_i \) for a base model \( f_i \) is derived based on its predictive accuracy \( S_i \) on the validation set. Assuming a Gamma distribution parameterised by shape \( \alpha \) and scale \( \beta \), the weights are given by:

\begin{equation}\label{eq:gprior}
w_i = \frac{S_i^{\alpha - 1} e^{-\frac{S_i}{\beta}}}{\sum_{j=1}^N S_j^{\alpha - 1} e^{-\frac{S_j}{\beta}}},
\end{equation}

\noindent where \( \alpha \) controls the shape of the prior distribution, \( \beta \) scales the prior weights, and \( S_i \) is the performance metric (e.g., accuracy or F1-score) of the \( i \)-th base model.

Updating the prior weight further strengthens the justification for this approach by ensuring that the model weights evolve based on empirical evidence. This is achieved by combining the prior weights with normalised likelihoods derived from model performance metrics.
The updated weights, or posterior weights, reflect both the initial beliefs and the observed performance, ensuring that the ensemble adapts dynamically to the strengths and weaknesses of the baseline models.
This dynamic weighting mechanism contrasts sharply with traditional ensembles, where weights are often static or determined heuristically.
After initialising prior weights using the Gamma distribution, the framework refines these weights by incorporating evidence from model performance. 
The likelihood of each model is computed as the normalised performance score as given in Equation~\eqref{eq:glikhood}:

\begin{equation}\label{eq:glikhood}
\text{Likelihood}(S_i) = \frac{S_i}{\sum_{j=1}^N S_j},
\end{equation}

\noindent This likelihood captures how well each model performs relative to the others. The prior weights are then updated using Bayes’ theorem, resulting in posterior weights as expressed in Equation~\eqref{eq:bupdate}:

\begin{equation}\label{eq:bupdate}
w_i^{\text{posterior}} = \frac{w_i^{\text{prior}} \cdot \text{Likelihood}(S_i)}{\sum_{j=1}^N w_j^{\text{prior}} \cdot \text{Likelihood}(S_j)}.
\end{equation}

\noindent Bayesian updating allows the ensemble to adapt to new information, which is crucial in multiclass classification scenarios where the effectiveness of the model can vary between different classes. Additionally, weight regularisation introduces another layer of robustness to Bayesian ensemble learning by addressing the risk of over-reliance on a single model. 
In multiclass tasks, where certain classes may be more challenging to distinguish, unregularised weights could lead to overfitting. This occurs when the ensemble becomes overly influenced by a model that performs well on specific classes but is poorly overall. Regularisation helps mitigate this by smoothing the weight distribution, ensuring balanced contributions from all models and enhancing the ensemble's ability to generalise across different classes. 
Here, regularised weights are computed as Equation~\eqref{eq:regularised}:

\begin{equation}\label{eq:regularised}
w_i^{\text{regularised}} = \frac{w_i^{\text{posterior}} + \lambda}{1 + \lambda \cdot N},
\end{equation}

\noindent where \( \lambda \) is the regularisation parameter, and \( N \) is the number of models. 
The shared weighting mechanism serves as the foundation for the different Bayesian ensemble strategies. 

Bayesian Averaging computes the ensemble prediction as a weighted average of the base model predictions to get Equation~\eqref{eq:BAEn}:

\begin{equation}\label{eq:BAEn}
P(y = c | x) = \sum_{i=1}^N w_i^{\text{regularised}} P_i(y = c | x),
\end{equation}

\noindent where \( P_i(y = c | x) \) represents the predicted probability distribution for class \( c \) by the \( i \)-th model. This approach enhances resilience to poorly performing models, as their contributions are down-weighted. It is particularly effective when the base models are diverse, as the averaging process leverages their complementary strengths.

For Bayesian Boosting, while using the same regularized weights, the focus is on iteratively improving performance by emphasising harder-to-classify instances. The ensemble prediction in Bayesian Boosting follows a similar formulation but introduces additional weighting during the prediction process as given in Equation~\eqref{eq:BAboost}:

\begin{equation}\label{eq:BAboost}
P(y = c | x) = \frac{\sum_{i=1}^N w_i^{\text{regularised}} P_i(y = c | x)}{\sum_{i=1}^N w_i^{\text{regularised}}}.
\end{equation}

\noindent  This formulation ensures that the final probabilities are normalized, maintaining valid distributions. By emphasising instances that are frequently misclassified, Bayesian boosting improves the ensemble’s ability to handle challenging cases, particularly in multiclass scenarios where class overlaps are common.

Bayesian Bagging applies the shared weights to predictions from models trained on bootstrap samples. The ensemble prediction is computed as Equation~\eqref{eq:BAEn}. 
This method leverages diversity introduced through bootstrap sampling, with the regularised weights ensuring robustness to noisy data and outliers. This combination of sampling diversity and probabilistic weighting makes Bayesian bagging well-suited for multiclass classification tasks with imbalanced datasets.

The final ensemble--Bayesian Stacking extends the framework by introducing a meta-learning layer. Base model predictions are transformed into meta-features, which are then used to train a meta-learner. 
Gaussian noise is added to the predictions to enhance diversity as depicted in Equation~\eqref{eq:BAnoise}:

\begin{equation}\label{eq:BAnoise}
\tilde{P}_i(y = c | x) = P_i(y = c | x) + \mathcal{N}(0, \sigma^2),
\end{equation}

\noindent where \( \mathcal{N}(0, \sigma^2) \) represents Gaussian noise with mean $0$ and variance \( \sigma^2 \). The meta-learner combines these noisy predictions using the regularised weights expressed in Equation~\eqref{eq:BAstack}:

\begin{equation}\label{eq:BAstack}
P(y = c | x) = g\left(\sum_{i=1}^N w_i^{\text{regularized}} \tilde{P}_i(y = c | x)\right),
\end{equation}

\noindent where \( g \) denotes the meta-learner. This hierarchical approach leverages the strengths of individual models while introducing an additional layer of optimisation, making it particularly powerful for resolving complex class boundaries.

 \section{Performance Metrics} \label{sec:pmetric}
In a multi-class classification scenario, such as a problem involving the categorisation of four distinct types of radio galaxies, the confusion matrix is generalised to a $4 \times 4$ table. Each row in the matrix corresponds to the actual class, and each column corresponds to the predicted class. The entries of this matrix provide a detailed breakdown of the classification performance for each class. Specifically, the diagonal elements represent the correctly classified instances for each class, referred to as the true positives (\(TP_{Ci}\)), while the off-diagonal elements capture instances where misclassifications occur. The general structure of the confusion matrix for a four-class classification problem is shown in Table~\ref{tab:layout}:

\begin{table*}[h!]
\centering
\caption{Confusion Matrix for Multi-Class Classification}\label{tab:layout}
\begin{tabular}{|c|c|c|c|c|}
\hline
            & \textbf{Predicted: C0} & \textbf{Predicted: C1} & \textbf{Predicted: C2} & \textbf{Predicted: C3} \\
\hline
\textbf{Actual: C0} & \( TP_{C0} \) & \( FP_{C0C1} \) & \( FP_{C1C3} \) & \( FP_{C0C3} \) \\
\textbf{Actual: C1} & \( FP_{C1C0} \) & \( TP_{C1} \) & \( FP_{C2C3} \) & \( FP_{C1C3} \) \\
\textbf{Actual: C2} & \( FP_{C2C0} \) & \( FP_{C2C1} \) & \( TP_{C2} \) & \( FP_{C2C3} \) \\
\textbf{Actual: C3} & \( FP_{C3C0} \) & \( FP_{C3C1} \) & \( FP_{C3C2} \) & \( TP_{C3} \) \\
\hline
\end{tabular}
\end{table*}

Here, \(TP_{Ci}\) represents the count of true positives for class \(C_i\), while \(FP_{CiCj}\) denotes instances of class \(C_i\) that were incorrectly classified as class \(C_i\). This matrix structure provides a foundation for computing a variety of performance metrics that evaluate the model's accuracy, precision, recall, and other measures across all classes.
For a given class \(C_i\), the performance metrics are defined as follows: True Positives (TP) are instances of \(C_i\) correctly classified as \(C_i\); False Positives (FP) are instances of other classes incorrectly classified as \(C_i\); False Negatives (FN) are instances of \(C_i\) misclassified as other classes; and True Negatives (TN) are instances of all other classes correctly not classified as \(C_i\).

Using these definitions, the common metrics are extended to multi-class classification as outlined in Table~\ref{table:metrics}:

\begin{table*}[h!]
\centering
\caption{Performance Metrics for Multi-Class Classification} \label{tab:metrics}
\begin{tabular}{@{}lll@{}}
\toprule
\textbf{Metric}       & \textbf{Description}                                                                 & \textbf{Computation}                                                      \\ \midrule
Precision (\(\text{Precision}_{{C_i}}\)) & Proportion of correct predictions for class \(C_{i}\).                            & \(\dfrac{TP_{C_i}}{TP_{C_i} + \sum_{j \neq i} FP_{C_jC_i}}\)               \\
& \\
Recall (\(\text{Recall}_{{C_i}}\))       & Proportion of actual instances of \(C_i\) correctly identified.                 & \(\dfrac{TP_{C_i}}{TP_{C_i} + \sum_{j \neq i} FP_{C_iC_j}}\)              \\
& \\
F1 Score (\(F1_{C_i}\))                & Harmonic mean of precision and recall for \(C_i\).                              & \(2 \cdot \dfrac{\text{Precision}_{C_i} \cdot \text{Recall}_{C_i}}{\text{Precision}_{C_i} + \text{Recall}_{C_i}}\) \\
& \\
Overall Accuracy                       & Proportion of all instances correctly classified.                               & \(\dfrac{\sum_{i=0}^{3} TP_{C_i}}{\text{Total Instances}}\)               \\
& \\
Macro-Averaged Metrics                 & Average performance across all classes.                                         & \(\dfrac{1}{4} \sum_{i=0}^{3} \text{Precision}_{C_i}\)                       \\
& \\
Micro-Averaged Metrics                 & Weighted average performance across all classes.                                & \(\dfrac{\sum_{i=0}^{3} TP_{C_i}}{\sum_{i=0}^{3} (TP_{C_i} + FP_{C_i})}\) \\ \bottomrule
\end{tabular}

\label{table:metrics}
\end{table*}

These metrics, particularly in the context of multi-class classification, are extensively discussed in \cite{sokolova2009systematic}. Their study highlights the importance of choosing the appropriate metric depending on the application and the relative significance of precision, recall, and overall accuracy.

\section{Results and Discussion} \label{sec:RnD}

\begin{figure*}
\begin{minipage}[H]{\textwidth}
\centering
\includegraphics[width=\textwidth]{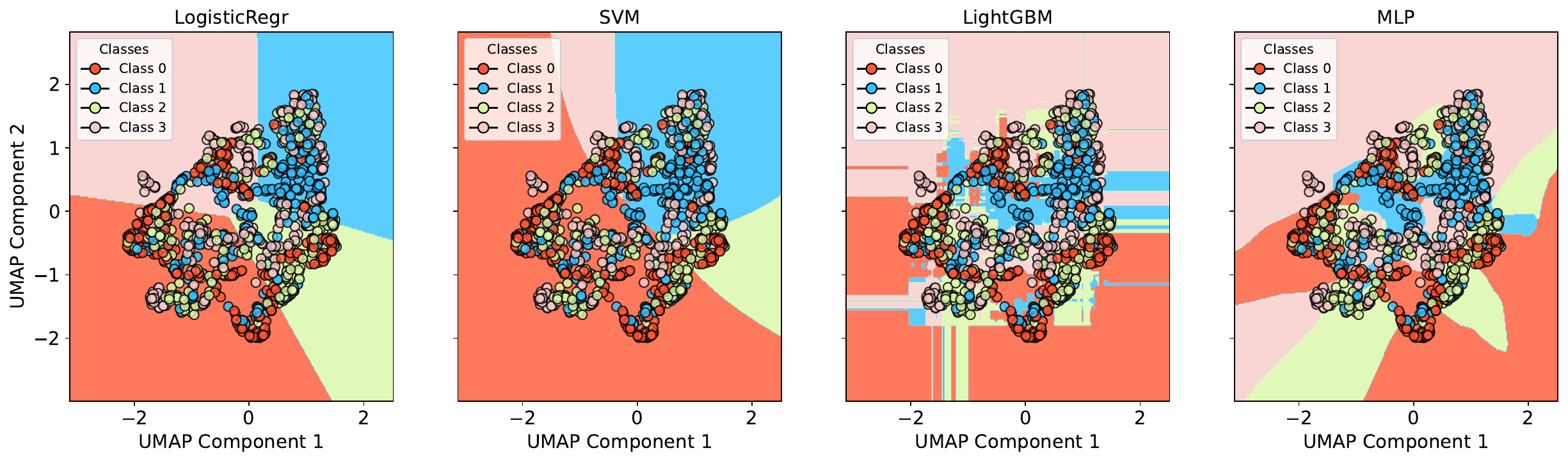} 
\end{minipage}
\caption{Decision Boundaries of Baseline Models on UMAP-transformed Radio Galaxy Data: This figure shows the decision boundaries of four baseline models--LogisticRegr, SVM, LightGBM, and MLP--trained on UMAP-transformed radio galaxy data. UMAP reduces the data's dimensionality while preserving its structure. Each plot overlays the model's decision boundaries on the UMAP-transformed data points, with colors indicating the predicted classes: 0 (Compact), 1 (Bent), 2 (FR-I), and 3 (FR-II). The plots visually compare how well each model separates the different classes in the feature space.
 }\label{fig:decision_boundaries}
\end{figure*}

In our multiclass classification scenario for enhanced clarity: \enquote{0} is assigned to the radio galaxy type Compact, \enquote{1} signifies the Bent radio galaxy category, \enquote{2} corresponds to the FR-I class, and \enquote{3} stands for the FR-II class. These categories were chosen to represent the primary types of radio galaxies based on their morphological and spectral properties.
To further analyse model performance, we examine the decision boundary plots in Fig.~\ref{fig:decision_boundaries}, which provide a visual representation of how each baseline model partitions the UMAP\footnote{Uniform Manifold Approximation and Projection}-transformed radio galaxy dataset into its predicted classes. These plots allow us to assess the models' ability to separate different classes and highlight potential areas of overlap or misclassification. 
UMAP, a method for reducing dimensionality while maintaining the manifold structure of the dataset, was utilised to simplify the complex feature space into a two-dimensional representation. This reduction is essential for visualising decision boundaries as it facilitates an intuitive depiction of the data and preserves key relationships among data points. By adeptly capturing both local and global structures within the data manifold, UMAP offers a more effective visualisation of decision boundaries than traditional linear reduction techniques.
The LogisticRegr model exhibits relatively simple linear decision boundaries, suggesting that it struggles to capture the complex nonlinear relationships inherent in the radio galaxy data. This is consistent with the lower accuracy observed in the previous analysis. 
The linear nature of the boundaries indicates that LogisticRegr primarily relies on linear combinations of features for classification, which may be insufficient to differentiate between the diverse morphological and structural characteristics of radio galaxies.
The SVM model, in contrast, demonstrates more flexible decision boundaries, capable of capturing nonlinear relationships in the data. The boundaries are smoother and more intricate than those of LR, suggesting a better ability to separate the classes in the feature space. This is consistent with its higher accuracy compared to LogisticRegr.
SVM's ability to handle nonlinearity through kernel functions likely contributes to its improved performance.
LightGBM, the highest-performing model, exhibits the most complex and adaptive decision boundaries. The boundaries are highly non-linear, with intricate shapes and variations across the feature space. This suggests that LightGBM effectively captures the intricate relationships between features, enabling it to distinguish between the subtle morphological and structural differences among the radio galaxy classes. The tree-based nature of LightGBM, which allows it to learn complex decision rules, likely contributes to its superior performance and the intricate nature of its decision boundaries.
The MLP model also demonstrates flexible decision boundaries, although they appear less intricate than those of LightGBM. The boundaries are somewhat smoother and less fragmented, indicating that MLP may be less sensitive to local variations in the data compared to LightGBM. This could be attributed to the MLP's reliance on global feature representations, which may be less effective in capturing fine-grained local patterns.
In our application of UMAP for dimensionality reduction and visualisation, the parameters were set as follows: n\_neighbors = 15, min\_dist = 0.1, and Euclidean distance as the metric. The UMAP embedding was applied to the balanced dataset with shape (1980, 120) obtained after the Local Neighbourhood Encoding process, rather than the original raw data. This approach ensures that the visualisation accurately reflects the features used for model training and classification, accounting for the class balancing step.

\begin{figure}
\begin{minipage}[H]{\linewidth}
\centering
\includegraphics[width=\textwidth]{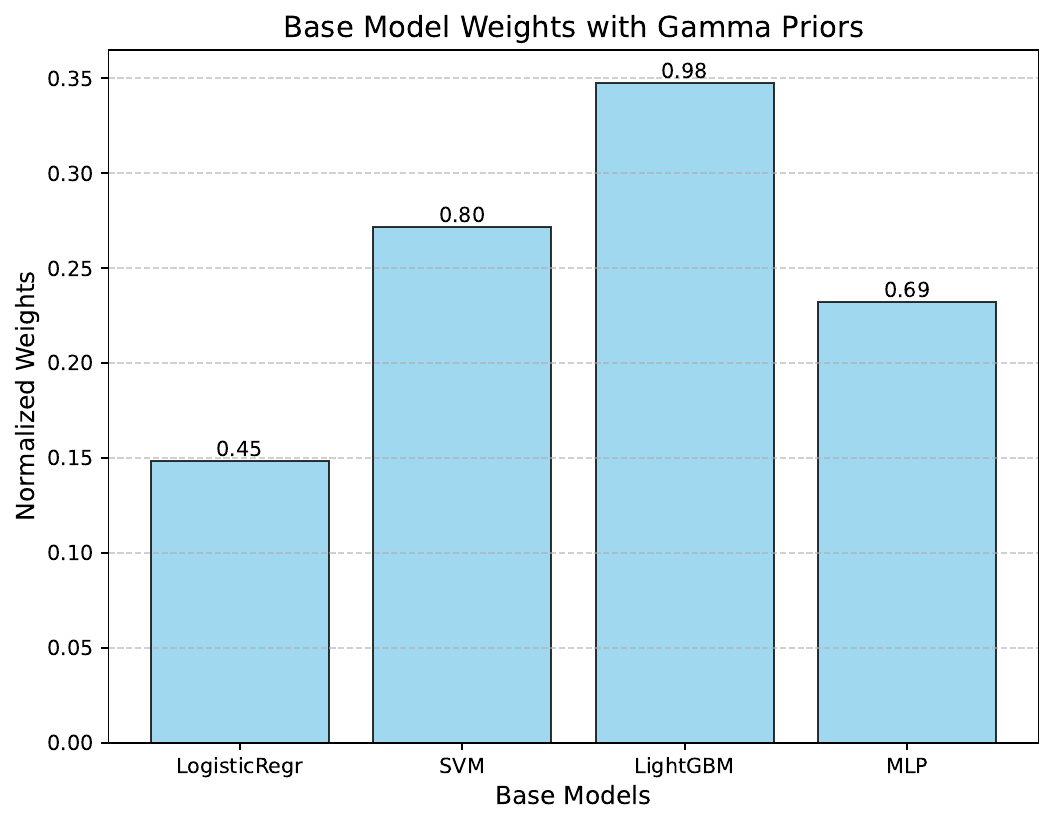} 
\end{minipage}
\caption{Performance and regularised weights of baseline models used for galaxy classification.
The bar heights represent the assigned regularised weights for each model in the Bayesian ensemble, while the numeric values on each bar indicate the classification accuracy of the respective models.
 }\label{fig:base_model_performance}
\end{figure}
The bar plots displayed in Fig.~\ref{fig:base_model_performance} provide a comprehensive visualisation of the performance metrics (accuracy) and regularized weights assigned to the four baseline models employed for galaxy classification. 
The Bayesian ensemble framework dynamically assigns weights to these models based on their accuracy while incorporating a regularisation parameter to ensure balanced contributions across the ensemble.
LogisticRegr exhibits the lowest accuracy of \(0.45\) among the four models, which is reflected in its relatively low regularized weight of $0.15$. 
This indicates limited predictive capability for this multinomial classification task.
LogisticRegr’s inability to effectively model the complex decision boundaries needed for distinguishing between diverse galaxy types is evident from its low performance. 
The reduced weight ensures that its influence in the Bayesian ensemble is minimal, preventing potential degradation of the ensemble’s overall performance.
The SVM model demonstrates a notable improvement in classification performance, achieving an accuracy of \(0.80\), accompanied by a higher regularised weight of \(0.27\).
This reflects the model's capability to handle the non-linear relationships present in the dataset. The higher accuracy suggests that SVM effectively separates the galaxy types by finding optimal hyperplanes, making it a critical contributor to the ensemble’s performance.
LightGBM emerges as the best-performing baseline model, achieving the highest accuracy of $0.98$, which justifies its assignment of the highest regularized weight of $0.35$.
This exceptional performance highlights LightGBM’s ability to capture complex patterns and interactions within the features, critical for classifying the distinct morphological and structural differences among galaxy types. The high weight ensures LightGBM plays a dominant role in the ensemble, significantly influencing the classification results.
The MLP model achieves an intermediate accuracy of $0.69$, with a corresponding regularised weight of $0.22$. 
While its performance is better than LogisticRegr, it falls short of SVM and LightGBM. This could be attributed to the MLP’s sensitivity to hyperparameter tuning and potential 
overfitting due to the dataset's characteristics. The regularised weight reflects a balanced contribution to the ensemble, allowing the model to complement stronger performers without overwhelming the ensemble’s decisions.

\begin{figure*}
\begin{minipage}[H]{\textwidth}
\centering
\includegraphics[width=\textwidth]{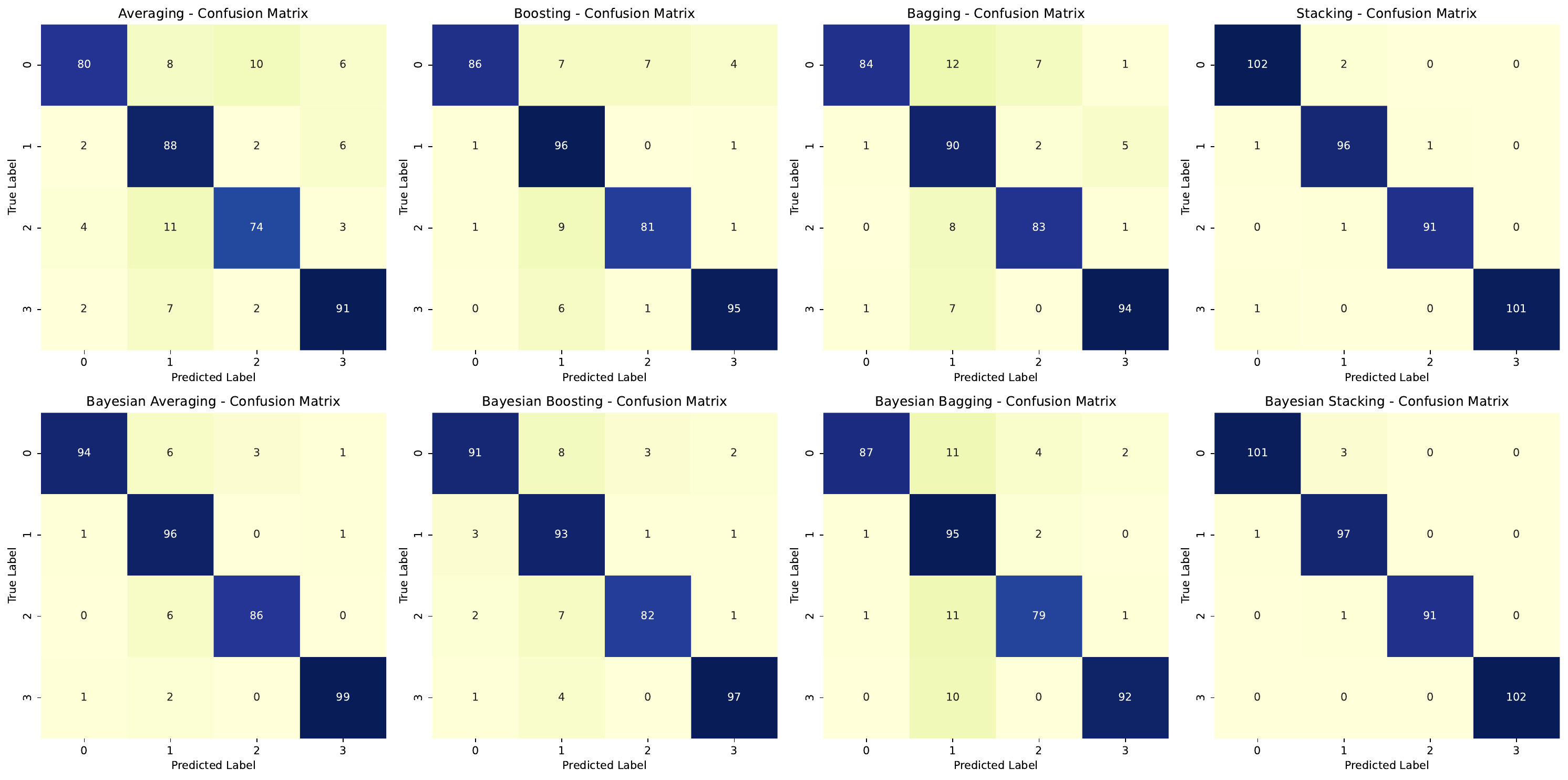} 
\end{minipage}
\caption{Confusion matrices for traditional and Bayesian ensemble methods in radio galaxy classification. The plots present the classification performance of four traditional ensemble methods (Averaging, Boosting, Bagging, and Stacking) in the first row and their Bayesian counterparts (Bayesian Averaging, Bayesian Boosting, Bayesian Bagging, and Bayesian Stacking) in the second row. Each matrix shows the number of correct and misclassified instances across four categories of radio galaxies: Compact (0), Bent (1), FR-I (2), and FR-II (3). The values along the diagonal represent the correct classifications, while off-diagonal values indicate misclassifications. The Bayesian ensemble methods exhibit superior performance in reducing misclassifications, particularly between the Compact and Bent categories, highlighting the advantages of incorporating Bayesian principles in ensemble learning for more accurate and generalized radio galaxy classification.
 }\label{fig:confusion_matrices}
\end{figure*}

The assessment of ensemble methods for classifying radio galaxies reveals notable performance disparities between conventional techniques and Bayesian-enhanced models, with Fig.~\ref{fig:confusion_matrices} providing insights into the precision, accuracy, and generalisation of each approach. The Averaging Ensemble, a conventional ensemble method, shows a fair but limited ability to differentiate between the classes. The model accurately identifies Compact galaxies in $80$ cases and Bent galaxies in $88$ cases, maintaining a satisfactory accuracy rate for FR-I ($71$) and FR-II ($91$). 
Nonetheless, appreciable misclassifications occur: particularly, $10$ Compact galaxies are misclassified as FR-I, and $8$ Bent galaxies are incorrectly identified as Compact. These findings indicate that although the Averaging Ensemble generally performs adequately, it faces challenges with distinguishing the subtle differences between the Compact and FR-I categories. This limitation could be due to the averaging mechanism, which, despite lowering variance, may not adequately represent the complexity involved in such detailed classification tasks, leading to a less precise decision boundary between these similar galaxy classes.

The Boosting Ensemble shows significantly enhanced performance, correctly classifying $86$ instances of Compact galaxies and $96$ instances of Bent galaxies. 
This model effectively addresses better performance across class distributions, especially for the Bent class. Misclassifications are minimal: only $1$ Bent galaxies are incorrectly labelled as Compact, and just $1$ for FR-II.  
Also,  FR-II category exhibits a low misclassification, a score of $1$ and $6$ for FR-I and Bent, respectively.
Boosting's iterative technique, which reweights the hardest-to-classify instances, allows for a nuanced decision-making process. As a result, the model excels at distinguishing between similar classes, such as Compact and Bent, thus enhancing overall classification accuracy. On the other hand, the Bagging Ensemble, while comparable to Boosting in general, shows a less substantial improvement in separating Compact from Bent galaxies. The confusion matrix shows $12$ Compact galaxies misclassified as Bent and $7$ Bent galaxies mislabeled as Compact. This suggests that although Bagging is effective against over-fitting and variability, it may not adequately address the intricate relationships between these classes. The model performs well with FR-I and FR-II galaxies, correctly classifying $94$ instances of the latter. However, its failure to sufficiently reduce mis-classifications in the more complex Compact and Bent categories results in slightly lower performance compared to Boosting.
The Stacking Ensemble, which combines the strengths of multiple base learners, achieves the highest accuracy among the traditional methods. 
The model performs exceptionally well across all four categories, with $102$ correct classifications for Compact galaxies and $96$ for Bent galaxies. This method excels in distinguishing between the different classes, particularly in the FR-I and FR-II categories, where it classifies $91$ and $101$ instances correctly, respectively. 
The relatively low rate of mis-classifications further highlights the potential of the Stacking method in tackling the challenges posed by the complex classification task. The stacking technique benefits from the diversity of its base models, allowing it to leverage complementary strengths in feature extraction and decision-making. 
As a result, Stacking exhibits superior performance in comparison to other traditional ensemble methods, making it a promising candidate for radio galaxy classification tasks.

Turning to the Bayesian Averaging Ensemble (in the second row of confusion matrices), this method offers a clear improvement over the traditional Averaging technique. The integration of Bayesian principles enhances the model's ability to generalise, as evidenced by the increased number of correct classifications for Compact and Bent galaxies ($94$ and $96$, respectively).
Mis-classifications are fewer, with only $6$ Bent galaxies being misclassified as Compact, and the performance for FR-I and FR-II categories is similarly improved. 
The Bayesian Averaging model’s improved performance can be attributed to the use of probabilistic modeling, which allows the ensemble to better manage uncertainty and make more robust predictions. 
The model’s ability to reduce mis-classifications and improve generalization is indicative of the advantages offered by Bayesian methods in ensemble learning.
The Bayesian Boosting Ensemble further refines the traditional Boosting approach, delivering significant improvements across all classes. Compact  and Bent galaxies benefit from reduced mis-classification rates, with only $8$ instances of Compact galaxies being misclassified and $3$ instances of Bent galaxies misclassified. The performance for FR-I and FR-II galaxies is similarly enhanced, with misclassification rates dropping to near zero for both categories. 
The Bayesian Boosting model’s success lies in its ability to better capture the underlying uncertainty in the data through probabilistic reasoning. By adjusting the weights of misclassified instances with greater precision, the Bayesian method improves the model’s focus on difficult-to-classify instances, leading to enhanced overall accuracy and a more refined decision boundary between the different classes.
The Bayesian Bagging Ensemble also shows notable improvements over its traditional counterpart, particularly in the classification of FR-II galaxies. With $92$ correct classifications for FR-II galaxies and a significant reduction in misclassifications between Compact and Bent galaxies, the Bayesian Bagging model achieves a higher level of precision. Mis-classifications for Compact and Bent galaxies drop to just 11 instances each, while the FR-I category also experiences enhanced performance. The Bayesian Bagging ensemble's probabilistic adjustments improve its decision-making, offering better generalisation and a more robust classification model, particularly for the more challenging galaxy types.
Finally, the Bayesian Stacking Ensemble emerges as the most effective model, achieving near-perfect classification across all four categories. The model classifies $102$ FR-II galaxies correctly, and the mis-classification rate for FR-I galaxies is virtually nonexistent. The performance improvements of the Bayesian Stacking model are attributed to its ability to combine the strengths of multiple base learners while incorporating Bayesian uncertainty quantification. This results in a model that is not only highly accurate but also capable of minimising mis-classifications between the various radio galaxy types. The Bayesian Stacking Ensemble’s performance underscores the potential of probabilistic ensemble learning methods in complex classification tasks, making it the most powerful tool for radio galaxy classification in this study.

\begin{figure*}
\begin{minipage}[H]{\textwidth}
\centering
\includegraphics[width=1.02\textwidth]{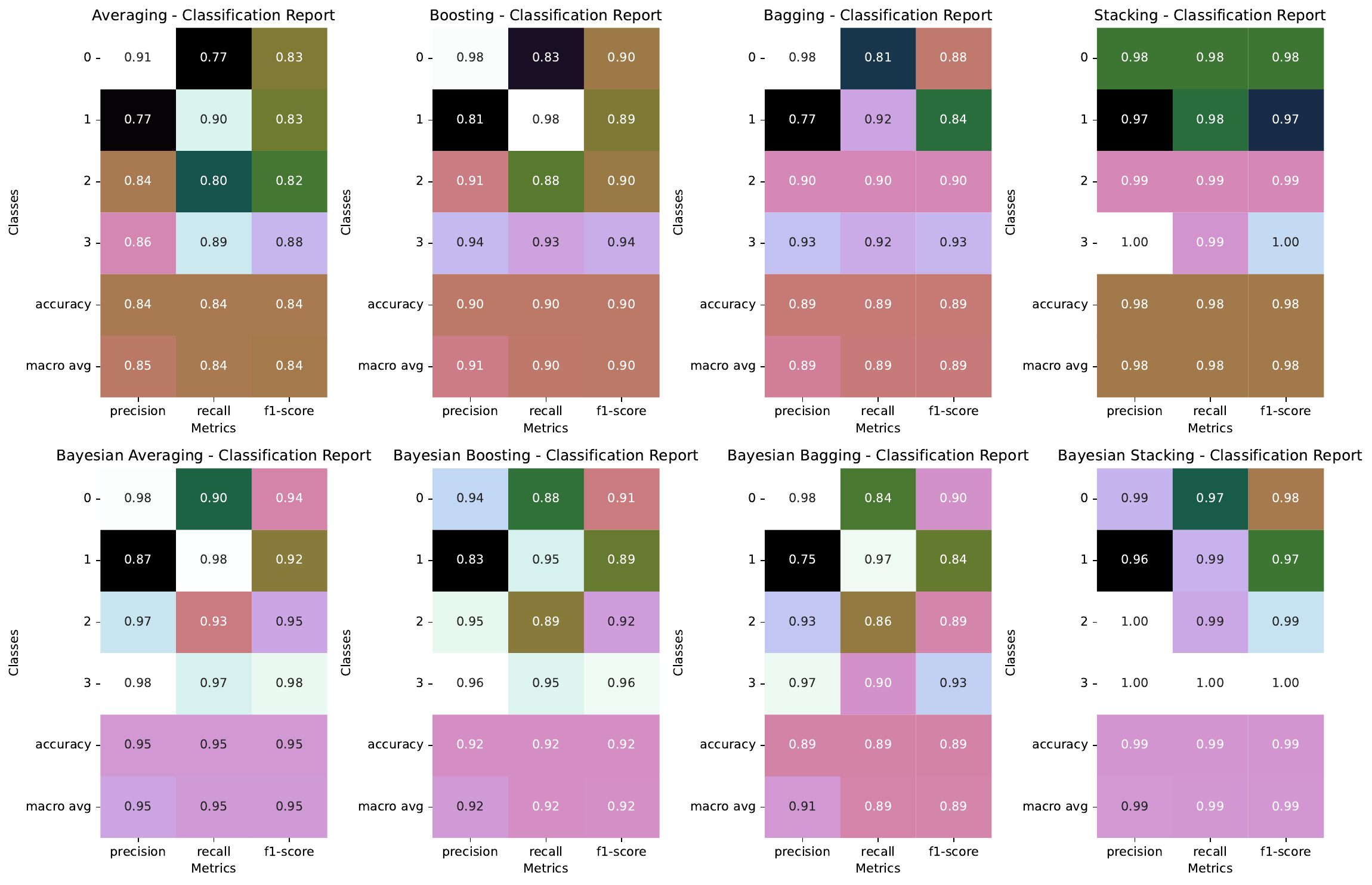} 
\end{minipage}
\caption{Classification performance reports of traditional and Bayesian ensemble models for radio galaxy classification. The plots show precision, recall, and F1-score for each class (Compact, Bent, FR-I, and FR-II) across four traditional ensemble methods (Averaging, Boosting, Bagging, and Stacking) and four Bayesian ensemble methods (Bayesian Averaging, Bayesian Boosting, Bayesian Bagging, and Bayesian Stacking). Additionally, accuracy, macro average, and weighted average scores are provided to evaluate the overall performance of each ensemble. The results highlight the superior classification capabilities of the Bayesian ensemble models, particularly the Bayesian Stacking ensemble, which achieves the highest accuracy and balanced metrics across all classes.
 }\label{fig:classification_reports}
\end{figure*}

Fig.~\ref{fig:classification_reports} provides a comprehensive assessment of the performance of both traditional and Bayesian ensemble approaches in the task of classifying radio galaxies.
Concerning traditional ensemble learning, the performance of the Averaging Ensemble is quite adept, achieving an overall accuracy of $84.0\%$. It exhibits high precision for Compact galaxies at $0.91$, though its recall is somewhat limited at $0.77$, suggesting some galaxies in this category are not identified.
Conversely, the Bent category shows a high recall of $0.90$, indicating effective detection of Bent galaxies, although the precision is slightly reduced to $0.77$. 
For the FR-I and FR-II categories, the Averaging method yields solid results, with F1-scores of $0.82$ and $0.88$, respectively, reflecting a well-maintained balance between precision and recall. However, the model’s difficulty in distinguishing Compact galaxies from other categories highlights the challenges in classifying similar types in this dataset.
The Boosting Ensemble outperforms the Averaging method, achieving a higher accuracy of $90.0\%$. Precision is notably high across all categories, particularly for Compact galaxies, where it reaches $0.98$. 
Although the recall for Compact galaxies is lower at $0.83$, the model compensates for this with a high F1-score of $0.90$.
Bent galaxies also benefit from Boosting, with precision at $0.81$ and recall at $0.98$, showing that the model is effective in detecting these galaxies while minimising false positives.
The Boosting model excels in classifying FR-I and FR-II galaxies, achieving F1-scores of $0.90$ and $0.94$, respectively. 
Boosting’s iterative error correction process proves to be a powerful tool, significantly enhancing classification accuracy, particularly for complex tasks like radio galaxy classification.
Achieving a solid accuracy of $89\%$, the Bagging Ensemble falls short of Boosting’s performance. Precision for Compact galaxies remains high at $0.98$, but recall drops to $0.81$, indicating a greater tendency to miss some of these galaxies. For Bent galaxies, both precision and recall are lower compared to Boosting, with F1-scores of $0.84$. When classifying FR-I and FR-II galaxies, Bagging performs similarly to Boosting, achieving F1-scores of $0.90$ and $0.93$, respectively. 
Bagging’s main strength lies in its robustness against variance, although it still faces difficulties in distinguishing between similar classes, such as Compact and Bent galaxies.
Among the traditional methods, the Stacking Ensemble achieves the highest accuracy, with an impressive score of $98\%$.
Precision and recall for all categories are outstanding, especially for FR-II galaxies, which are classified with near-perfect precision ($1.00$) and recall ($0.99$). 
The model achieves near-perfect F1-scores across all categories, with an overall macro average F1-score of $0.98$. This exceptional performance reflects the Stacking method’s ability to effectively combine the strengths of various base learners, making it particularly well-suited for classifying radio galaxies, where subtle distinctions between types are crucial.

In the case of Bayesian Ensemble Methods, the Bayesian Averaging Ensemble demonstrates significant improvement over its traditional counterpart, achieving an accuracy of $95.0\%$.
Precision and recall for Compact galaxies improve, with precision at $0.98$ and recall at $0.90$, indicating that the model is better at identifying these galaxies. 
Bent galaxies also benefit from higher precision at $0.87$ and near-perfect recall at $0.98$, showing the model's strong ability to classify this category while minimizing false positives.
The FR-I and FR-II categories see further enhancement, with F1-scores of $0.95$ and $0.98$, respectively. 
By incorporating probabilistic adjustments, the Bayesian Averaging method handles uncertainty more effectively, leading to more accurate overall classifications.
The Bayesian Boosting Ensemble takes performance even further, achieving an accuracy of $92.0\%$. Precision and recall for Compact galaxies improve to $0.94$ and $0.88$, respectively, although there is a slight decrease compared to the Bayesian Averaging method. Bent galaxies show precision and recall values of $0.83$ and $0.95$, respectively, with an F1-score of $0.89$, reflecting strong classification while maintaining balance. 
For FR-I and FR-II galaxies, performance remains robust, with F1-scores of $0.92$ and $0.96$, respectively. 
By integrating probabilistic insights, the Bayesian Boosting model refines the decision-making process, improving generalisation and handling imbalanced data more effectively.
The Bayesian Bagging Ensemble achieves an accuracy of $89.0\%$, showing notable improvements over the traditional Bagging method.
Precision for Compact galaxies is consistent at $0.98$, but recall increases to $0.84$, suggesting better identification of these galaxies. 
For Bent galaxies, recall improves significantly to $0.97$, although precision drops slightly to $0.75$. FR-I and FR-II categories benefit from strong precision ($0.93$ and $0.97$, respectively) and recall, resulting in good overall performance. The Bayesian Bagging method’s enhanced ability to manage uncertainty allows it to maintain a balance between precision and recall, which is crucial for classifying complex astrophysical data.
The Bayesian Stacking Ensemble delivers the highest accuracy of all the models, with an impressive $99.0\%$. Precision and recall for all categories are near-perfect, particularly for FR-II galaxies, which are classified with flawless precision ($1.00$) and recall ($1.00$). 
Bent and FR-I galaxies also exhibit excellent classification metrics, with F1-scores of $0.97$ and $0.99$, respectively. The Bayesian Stacking method’s ability to combine multiple base learners while incorporating Bayesian uncertainty results in a highly robust and accurate model. Its exceptional performance suggests that it is particularly well-suited for tasks requiring fine-grained classification, where minimising misclassifications is critical.

\begin{figure}
\begin{minipage}[H]{\linewidth}
\centering
\includegraphics[width=\textwidth]{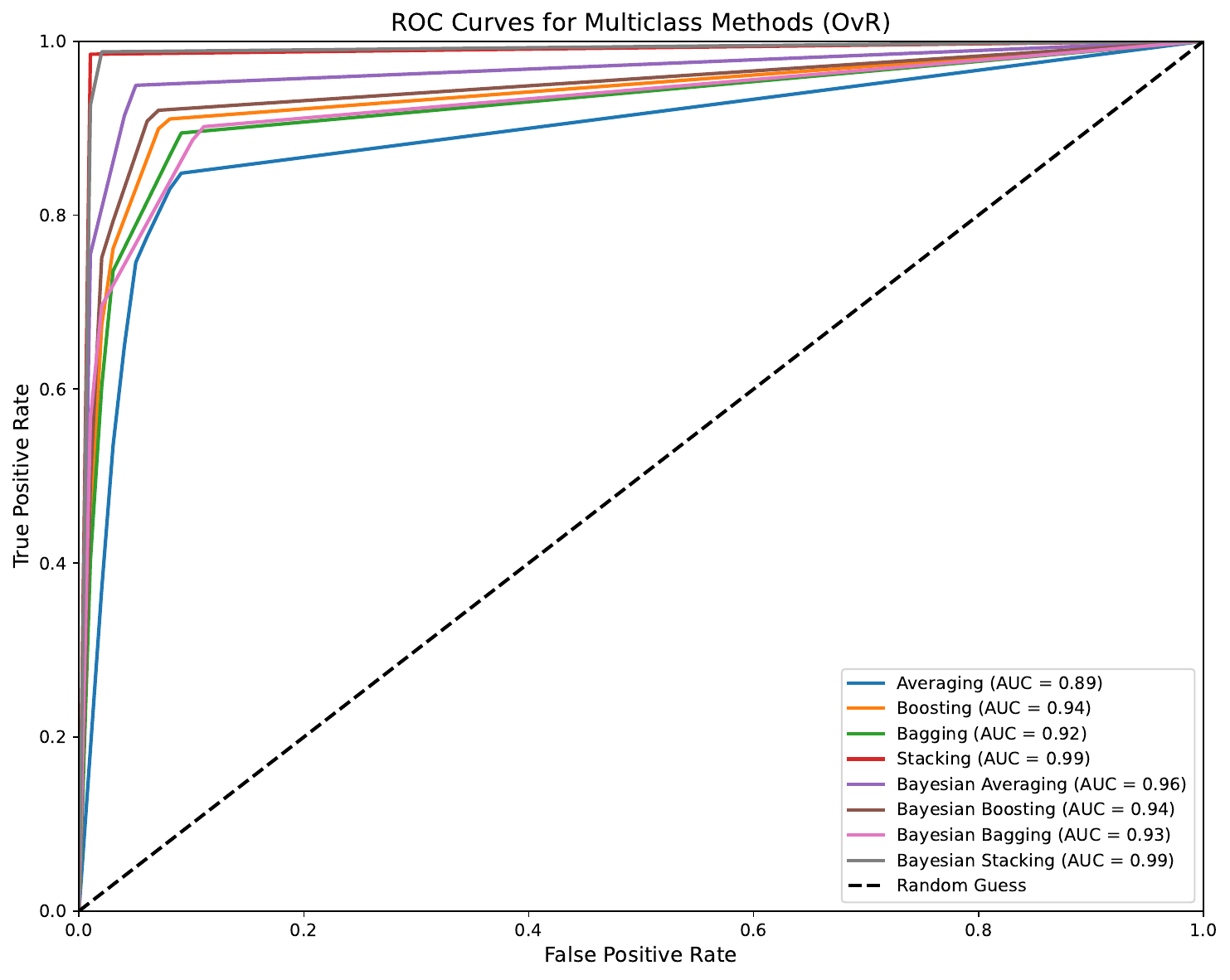} 
\end{minipage}
\caption{ROC curves for multiclass classification using traditional and Bayesian ensemble methods for radio galaxy classification. The plots depict the True Positive Rate (TPR) versus False Positive Rate (FPR) for each ensemble, with the Area Under the Curve (AUC) values included as a measure of classification performance. The Bayesian ensembles, particularly Bayesian Stacking ($\rm AUC = 0.99$), demonstrate superior separability among galaxy classes (Compact, Bent, FR-I, and FR-II), underscoring their effectiveness in addressing the challenges of morphological classification in radio astronomy.
 }\label{fig:roc_curves}
\end{figure}

The ROC\footnote{Receiver Operating Characteristic} curves presented in Fig.~\ref{fig:roc_curves} provide a comprehensive evaluation of the performance of traditional and Bayesian ensemble methods for multiclass classification of radio galaxies.
The analysis focuses on the ability of these models to discriminate among the classes (Compact, Bent, FR-I, and FR-II) using a one-vs-rest (OvR\footnote{In the OvR approach for multiclass classification, a separate classifier is trained for each class to distinguish it from all other classes.}) approach, with the AUC\footnote{Area Under the Curve} values serving as a critical metric for comparison.
The traditional ensemble methods--Averaging, Boosting, Bagging, and Stacking--exhibit varying degrees of classification performance, as evidenced by their AUC values. Among these, Stacking achieves the highest AUC of $0.99$, indicating excellent separability across classes. Boosting follows with an AUC of $0.94$, demonstrating strong classification capability, albeit slightly less robust than Stacking. Bagging, with an AUC of $0.92$, performs reasonably well but falls short of the top-performing traditional ensembles. Averaging, on the other hand, achieves an AUC of $0.89$, which, while indicative of fair performance, underscores its relative limitations in effectively capturing the nuances of the class distinctions.
In contrast, the Bayesian ensemble methods demonstrate superior performance overall, showcasing the benefits of integrating Bayesian principles into ensemble learning for this domain. Bayesian Stacking emerges as the best-performing model with an AUC of 0.99, matching the traditional Stacking ensemble but benefiting from the probabilistic robustness of the Bayesian framework. Bayesian Averaging achieves an impressive AUC of $0.96$, outperforming all traditional ensembles except for Stacking. Similarly, Bayesian Boosting and Bayesian Bagging attain AUC values of $0.94$ and $0.93$, respectively, illustrating their enhanced classification capabilities compared to their traditional counterparts.
The high AUC values across the board indicate that the ensembles effectively separate the galaxy classes, which is particularly crucial in the domain of radio astronomy. Accurate classification is vital for distinguishing among morphological types of radio galaxies, as these classes provide insights into the physical and dynamical processes governing galaxy evolution. The Bayesian Stacking ensemble's exceptional performance can be attributed to its ability to integrate the strengths of multiple classifiers while leveraging the Bayesian framework for better uncertainty modelling, a critical factor in astronomical datasets, which often suffer from inherent noise and imbalances.
Furthermore, the ROC curve demonstrates a notable improvement in classification performance with the Bayesian ensembles, particularly for challenging cases where inter-class overlaps exist. This improvement is a testament to the probabilistic aggregation of predictions, allowing Bayesian models to better handle ambiguities and uncertainties that arise in the classification process.

\begin{figure}
\begin{minipage}[H]{\linewidth}
\centering
\includegraphics[width=\textwidth]{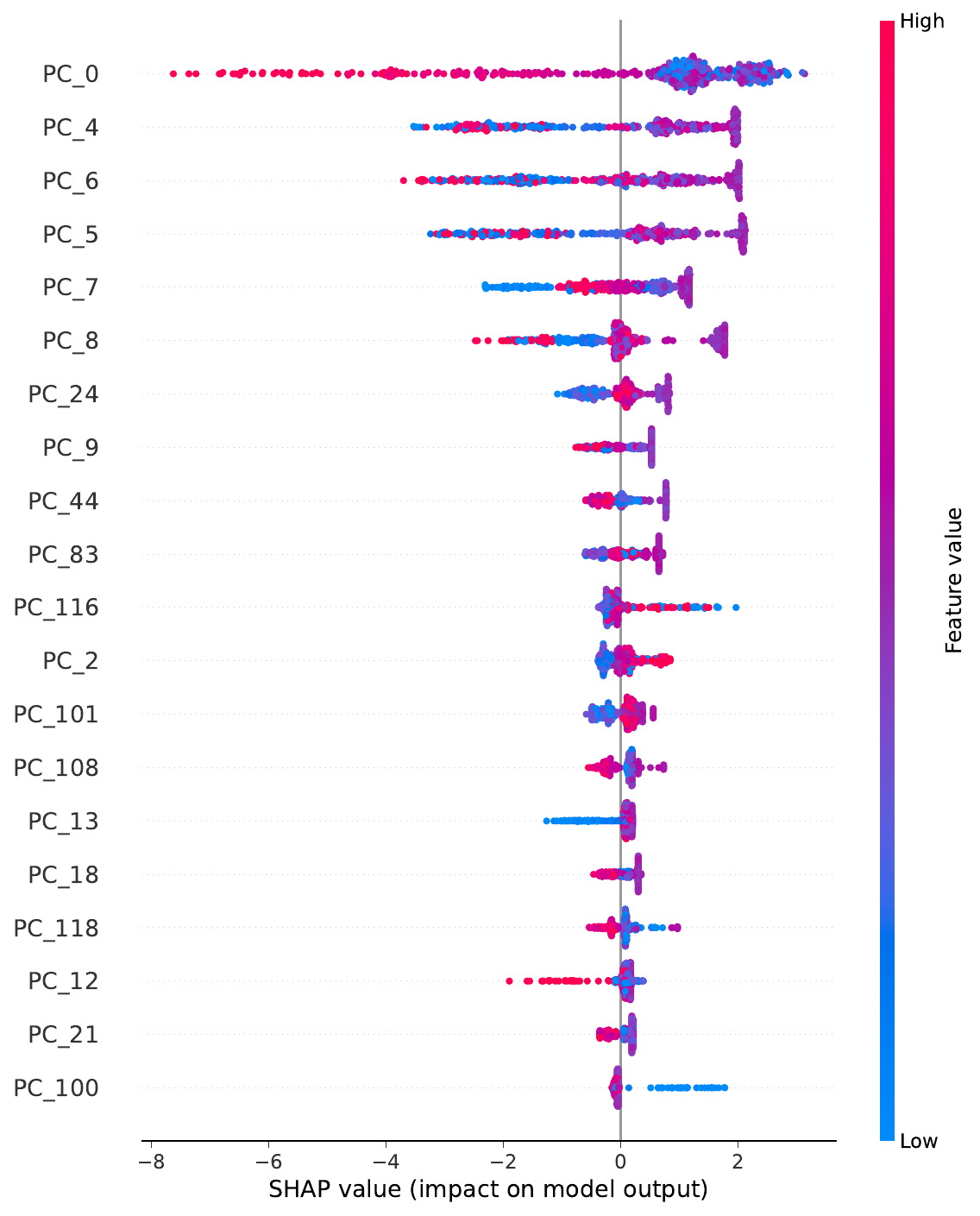} 
\end{minipage}
\caption{SHAP Values for LightGBM Model on SVD-based Features of Radio Galaxies: The plot displays the SHAP values for each principal component in the LightGBM model, which was trained using features extracted from the SVD of radio galaxy data. The x-axis represents the SHAP values, showing how each feature influences the model's output, with positive values indicating an increase and negative values a decrease. The y-axis lists the principal components, while the colour of each dot reflects the feature value. This visualisation helps to understand the contribution of each principal component to the model's classification decisions.
 }\label{fig:LightGBM_summary}
\end{figure}

The SHAP plot shown in Fig.~\ref{fig:LightGBM_summary} illustrates how each principal component (PC) derived from the SVD of radio galaxies influences the LightGBM model's predictions for each data point.
On the x-axis is the SHAP value, which measures a feature's influence on the model's prediction. Features with positive SHAP values contribute to increasing the model's output (i.e., they shift the prediction towards a higher class), whereas features with negative SHAP values decrease the model's output. The y-axis displays the various principal components, and the colour of each dot signifies the feature's value.
$\rm PC\_0,\, PC\_4,\, PC\_6,\, PC\_7,\, and\,  PC\_8$ consistently show high positive SHAP values, indicating a strong influence in driving the model's output upwards. This suggests that these PCs likely capture features related to specific types of radio galaxies, such as morphological characteristics (e.g., size, shape, jet orientation) or spectral properties (e.g., radio luminosity, spectral index), which are indicative of distinct classes.
$\rm PC\_2, \, PC\_101, \, PC\_108, \, PC\_13, \, PC\_18, \,  and \, PC\_118$ exhibit a combination of positive and negative SHAP values, reflecting that their impact on the model's output varies depending on the specific values. This implies that these PCs capture features that are not consistently linked to a single class, but rather may help differentiate between various types of radio galaxies based on specific combinations of characteristics.
The SHAP plot also reveals that $\rm PC\_24,\, PC\_44,\, PC\_83,\, and \, PC\_116$ exhibit low or negligible SHAP values--indicating that they have little influence on the model's predictions.
This observation provides valuable insights into the relative importance of different features in the classification of radio galaxies.
These PCs may capture features that are either irrelevant for classification, representing noise or variations in the data that do not help distinguish between different radio galaxy types, or redundant, with information already captured by other PCs with higher SHAP values. As a result, they would contribute little to the model's predictive power. However, it is important to note that the \enquote*{unimportance} of these PCs is relative. At the same time, they may have minimal impact on the current LightGBM model; they could potentially be more relevant for other classification models or tasks, such as anomaly detection.
The SHAP values presented in Fig.~\ref{fig:LightGBM_summary} were computed based on the model’s predictions for all 396 test samples in our dataset. The selection of LightGBM for SHAP analysis was driven by its superior predictive performance, as evidenced by the highest accuracy and AUC scores among the evaluated classifiers. Additionally, the tree-based structure of LightGBM facilitates efficient and interpretable computation of SHAP values, enabling us to gain detailed insights into feature contributions. Due to the computational complexity of calculating SHAP values for ensemble models with high dimensionality, focusing on the best-performing model allowed for a comprehensive and insightful interpretability analysis. In future studies, alternative models could be similarly interrogated, but LightGBM’s robustness and interpretability made it the most suitable choice for the current analysis.

A comparison between the singular value spectrum (Fig.~\ref{fig:_svdplot_}) and the SHAP-based component importance (Fig.~\ref{fig:LightGBM_summary}) reveals that variance strength and discriminative relevance are related but not identical. 
Components with the largest singular values capture dominant variance directions, while SHAP analysis identifies those components most influential for class separation. 
In our results, the leading components (PC\_1--PC\_4) appear prominently in both analyses, suggesting partial correspondence, yet several mid-rank components also exhibit high SHAP importance. 
This observation highlights that features explaining moderate variance can still encode strong class-discriminative structure, an effect consistent with previous findings in representation learning \citep{tharwat2016principal,shlens2014tutorial}.

\section{Conclusion} \label{sec:conc}

Our study has introduced a comprehensive framework that integrates Bayesian ensemble learning techniques with SVD-based feature extraction for automated radio galaxy classification. The methodologies presented address significant challenges in handling complex, noisy, and high-dimensional astronomical data, achieving improved classification accuracy and interpretability.

A critical component of our approach involved the LNE method to address class imbalance in the dataset. Unlike traditional techniques like SMOTE or under-sampling methods, which risk overpopulation or loss of valuable information, LNE preserved the intrinsic structure of the data. By generating synthetic samples within local neighbourhoods, LNE balanced the dataset while maintaining astrophysical validity, ensuring that the classifier was trained on a dataset representative of the underlying distributions. This step effectively eliminated bias in the classification task and established a solid foundation for subsequent analyses.  
The integration of Bayesian ensemble learning provided robust mechanisms for combining predictions from multiple classifiers. Among the models evaluated, LightGBM emerged as the most effective base learner, as evidenced by its consistently superior performance in decision boundary delineation and SHAP interpretability. Higher weights were assigned to LightGBM in the Bayesian ensemble, reinforcing its importance in identifying intricate patterns within the feature space. 
SHAP analysis further revealed that LightGBM was particularly adept at capturing key features and their relationships with specific radio galaxy types, highlighting its suitability for handling complex astrophysical data. 
The SVD employed for dimensionality reduction, proved instrumental in extracting PCs that retained the critical morphological and spectral characteristics of radio galaxies. High SHAP values associated with PCs such as PC\_0, PC\_4, PC\_6, PC\_7, and PC\_8 demonstrated their strong association with distinct radio galaxy classes. These PCs encapsulated features likely corresponding to fundamental morphological and spectral traits crucial for distinguishing between FR-I, FR-II, Compact, and Bent galaxy types. While these components have shown significant importance in classification, further astrophysical analysis is required to elucidate their physical meaning and relationship to observed galaxy properties. Future research could explore the underlying astrophysical significance of these features to provide deeper insights into radio galaxy evolution. 
The PSO algorithm played a vital role in optimising hyperparameters across various base models (LogisticRegr, SVM, LightGBM, and MLP), enhancing their performance and ensuring that the ensemble was fine-tuned for optimal accuracy. By leveraging PSO, the framework dynamically adapted the parameters of complex models such as LightGBM and multilayer perceptrons, maximising their contributions to the ensemble's overall predictive power. The inclusion of PSO underscores the importance of advanced optimisation techniques in developing scalable and high-performing machine learning frameworks for astronomical applications. 
Our findings underscore the potential of integrating SVD with ensemble learning approaches, particularly Bayesian methods, for addressing challenges in radio galaxy classification. The Bayesian stacking ensemble, which combined the outputs of optimised base models, emerged as the most effective in capturing uncertainty and delivering high classification accuracy. Importantly, the framework demonstrates that the combined use of advanced feature extraction and optimised ensemble methods can significantly enhance the reliability and interpretability of machine learning models in astronomy.

Beyond addressing immediate classification challenges, the proposed framework demonstrates broader applicability to large-scale astronomical data analysis. 
To evaluate its computational efficiency and scalability, the LNE-based ensemble pipeline was implemented and executed on a High Performance Computing (HPC) cluster. 
For a training set of approximately $10^{4}$ sources, a complete model run, including dimensionality reduction, feature selection, and nested cross-validation, required about $2.4$~hours on a configuration utilising two compute nodes, each with eight CPU cores at $3.6$~GHz and $32$~GB of RAM. 
This setup yielded an estimated energy footprint of approximately $0.22$~kWh per full training cycle, representing about an order-of-magnitude reduction in computational cost relative to comparable deep learning baselines reported for similar classification tasks. 
These performance characteristics highlight the framework’s scalability and adaptability, making it particularly well suited for integration into the data-processing pipelines of next-generation surveys such as the SKA and the Vera~Rubin~Observatory.

By tackling issues of high-dimensionality, noise, and class imbalance, this methodology sets a precedent for managing the data deluge anticipated in these projects. 
Moreover, the versatility of the approach suggests its applicability to a wide range of domains requiring automated classification of complex datasets. Examples include detecting transient phenomena, classifying stellar populations, or analysing gravitational wave signals. The fusion of advanced feature extraction, ensemble learning, and optimisation techniques exemplified in this study can inspire similar advancements across other data-intensive sciences.

Beyond the immediate technical contributions, this work offers several conceptual advancements with implications for both astronomy and ML. 
By demonstrating that Bayesian ensemble learning effectively addresses class imbalance and observational noise in radio galaxy datasets, our framework highlights a paradigm shift toward probabilistic, uncertainty-aware methodologies in astrophysical classification. 
This aligns with broader trends in data-driven science, where quantifying predictive confidence becomes critical for interpreting complex, high-dimensional datasets.
For instance, the ability to dynamically weight model contributions via Bayesian inference could inform strategies for next-generation surveys like the SKA, where automated pipelines must balance scalability with robustness to heterogeneous data quality.
Furthermore, the integration of SVD-based feature extraction underscores the importance of preserving morphological fidelity in dimensionality reduction, a principle that may extend to studies of galaxy evolution, where subtle structural variations encode clues about cosmic expansion or black hole dynamics. Such cross-disciplinary insights position our framework as a template for tackling similar challenges in transient detection, stellar population analysis, or gravitational wave signal classification.

While our framework demonstrates strong performance, it is important to acknowledge its limitations. The LNE method, though effective at addressing class imbalance by generating synthetic samples in the SVD-decomposed feature space, assumes that these artificially created instances maintain astrophysical validity. This assumption hinges on the SVD-derived principal components (PCs) adequately capturing the critical morphological and spectral characteristics of radio galaxies. However, for highly underrepresented classes, such as Compact or Bent galaxies, that exhibit sparse or nonuniform distributions in the feature space, LNE’s local interpolation may inadvertently smooth over or distort subtle structural variations encoded in the PCs. For example, if key PCs (e.g., PC\_24, PC\_44) retain low SHAP values and are dominated by noise rather than meaningful astrophysical signals, synthetic samples generated in this space could propagate artefacts rather than realistic representations. Furthermore, real-world deployment of the model is subject to concept drift that is, gradual shifts in data characteristics or classification criteria over time.
As new morphological types are discovered and taxonomies evolve, the model’s learned boundaries may become outdated, requiring ongoing monitoring, retraining, and recalibration to ensure continued reliability and relevance.

In summary, this work presents a framework that advances automated radio galaxy classification and lays the groundwork for future developments in astronomical data analysis. Further research should explore the astrophysical meaning of the extracted principal components and evaluate alternative feature extraction and classification techniques. Approaches such as variational autoencoders or graph-based learning may complement the current framework, offering deeper insights into the morphological diversity of radio galaxies. Additionally, domain adaptation strategies could improve model transferability across heterogeneous surveys (e.g., from FIRST to SKA), reducing biases introduced by instrument-specific noise or resolution disparities.
Promising future directions include investigating Dirichlet-weighted aggregation as an extension to Gamma-weighted methods. Unlike static priors, Dirichlet distributions model uncertainty across multiple contributors in multiclass settings, allowing dynamic weight adjustment during training. This could improve resilience to class imbalance and concept drift by aligning model focus with evolving data characteristics.
Integrating physics-informed constraints such as morphological symmetry or spectral energy distribution templates into ensemble decision boundaries would also improve interpretability, bridging machine learning with domain expertise.
Lastly, applying this pipeline to time-domain astronomy, such as classifying variable radio sources, would test its adaptability to shifting class distributions, reinforcing its utility in the era of big data astrophysics.

\section*{CRediT authorship contribution statement}
\textbf{Theophilus Ansah-Narh:} Conceived the study, designed the framework, and implemented Bayesian ensemble learning and SVD-based feature extraction for automated radio galaxy classification. \\
\textbf{Jordan Lontsi Tedongmo:} Developed the Local Neighborhood Encoding  methodology for addressing class imbalance and contributed to the data preprocessing pipeline. \\
\textbf{Joseph Bremang Tandoh:} Conducted performance evaluation of classifiers, including LightGBM and multilayer perceptrons, and led the optimisation process using Particle Swarm Optimisation. \\
\textbf{Nia Imara:} Provided critical insights and expertise in the interpretation of the radio survey datasets, particularly in understanding the astrophysical properties and morphological classifications derived from the FIRST, NVSS, and LoTSS surveys. Contributed to the discussion on data integration strategies, cross-matching procedures, and the implications of multi-frequency observations for radio galaxy classification.\\
\textbf{Ezekiel Nii Noye Nortey:} Provided insights on SHAP interpretability, contributed to the interpretation of principal components, and reviewed the manuscript for astrophysical validity.

\section*{Declaration of competing interest}
The authors declare that there are no contending financial interests or personal relationships that could give the impression of influencing this project.

\section*{Data availability} 

The dataset used in this study is a combination of several catalogues
characterising radio sources from the FIRST survey \citep{Becker1995}. It has been published by \cite{rustige2023morphological} and is available on \href{https://github.com/floriangriese/RadioGalaxyDataset}{GitHub}. The dataset used comprises $2156$ radio galaxy images with morphological labels, grouped into four classes: FR-I, FR-II, Compact and Bent.
The dataset used in this study is a combination of several catalogues characterising radio sources from prominent sky surveys. Primarily, it includes data from the FIRST survey \citep{Becker1995}, renowned for its high-resolution imaging at $1.4$ GHz, which has been extensively utilised in radio galaxy morphology studies. This dataset has been published and made publicly accessible by \cite{rustige2023morphological} on \href{https://github.com/floriangriese/RadioGalaxyDataset}{GitHub}, containing a total of $2,156$ radio galaxy images with annotated morphological labels. In addition to FIRST, the dataset incorporates sources from the NRAO VLA Sky Survey \citep{condon1998nrao}, which provides lower-resolution, wide area coverage at $1.4$ GHz, valuable for capturing large-scale structures and the LoTSS \citep{shimwell2017lofar}, offering low-frequency observations at 120--168 MHz that reveal extended diffuse emission features. The combined dataset encompasses images categorised into four classes: FR-I, FR-II, Compact, and Bent radio galaxies, facilitating robust morphological classification and analysis.

\section*{Funding}

This research was fully supported under the collaborative framework established by the \textit{Memorandum of Understanding} between the Ghana Atomic Energy Commission (GAEC), acting through the Ghana Space Science and Technology Institute (GSSTI), and the Regents of the University of California on behalf of its Santa Cruz Campus (UCSC). 
The collaboration focuses on advancing joint research and educational activities in astronomy and astrophysics, with particular emphasis on the development and scientific utilisation of the Ghana Radio Astronomy Observatory (GRAO). 
Funding and institutional support for this project were provided through the GRAO-UCSC Astronomy Development Project. 
The authors gratefully acknowledge the logistical and technical assistance of the GAEC/GSSTI and the UCSC Department of Astronomy and Astrophysics.

\section*{Acknowledgments}

The authors would like to express their gratitude to the anonymous referees for their insightful and constructive feedback. Additionally, TA-N extends sincere appreciation for the support provided by the Ghana Space Science and Technology Institute (GSSTI), as well as for the access to their High-Performance Computing (HPC) resources, which facilitated the completion of this work.

\bibliography{sample}

\end{document}